\documentclass[aps,pre,reprint,10pt,superscriptaddress,showpacs]{revtex4-2}
\usepackage{amsfonts} 
\usepackage{amsmath}
\usepackage{amssymb}
\usepackage{graphicx}
\usepackage{subfigure}
\usepackage{color}
\usepackage{color}
\usepackage{soul}
\usepackage{cancel}
\usepackage{ulem}
\newcommand*\xbar[1]{%
	\hbox{%
		\vbox{%
			\hrule height 0.5pt 
			\kern0.5ex
			\hbox{%
				\kern-0.1em
				\ensuremath{#1}%
				\kern-0.1em
			}%
		}%
	}%
} 

\begin{document}
	\title[]{Chaos in the dynamics of electromagnetic solitons in relativistic degenerate plasmas}
	\author{Subhrajit Roy}
	\email{suvo.math88@gmail.com}
	\affiliation{Department of Mathematics, Visva-Bharati University, Santiniketan-731 235, India}
	\author{S. Das Adhikary}
	\email{sukhendusda@gmail.com}
	\affiliation{Department of Mathematics, Visva-Bharati University, Santiniketan-731 235, India}
	\author{Amar P. Misra}
	\email{apmisra@visva-bharati.ac.in}
	\homepage{Author to whom any correspondence should be addressed.}
	\affiliation{Department of Mathematics, Visva-Bharati University, Santiniketan-731 235, India}
	
	\date{\today }
	\begin{abstract}
		We propose a coupled system for the nonlinear interaction between high-frequency, circularly polarized, intense electromagnetic (EM) waves and low-frequency electron-density perturbations, driven by the EM-wave ponderomotive force, in an unmagnetized plasma composed of fully degenerate relativistic electrons and stationary positive ions, including a higher-order correction to the nonlocal nonlinearity. We show that the modulational instability (MI) growth rate associated with the generation of EM envelope solitons gets significantly reduced with a slight increase in either the nonlocal nonlinear correction or the degeneracy parameter.  Furthermore, a three-wave temporal model predicts the existence of quasiperiodic and chaotic states of EM solitons while interacting with longitudinal electron density perturbations. We show that the greater the degeneracy (or higher the contribution from the nonlocal correction), the smaller the instability domain of modulation wave numbers; thus, degeneracy favors the stability of EM soliton evolution. The existence of temporal chaos in a low-dimensional model could be a signature of the development of spatiotemporal chaos in the complete nonlinear model, in which many electromagnetic solitons can be excited and saturated as they interact with electron plasma waves.     
	\end{abstract}
	\maketitle
	\textbf{The nonlinear interactions between electromagnetic (EM) waves and electrostatic density perturbations are fundamental in generating electromagnetic solitons via modulational instability and in the onset of chaos in space and astrophysical environments. Such interactions are also helpful for particle heating, energy transport, and turbulence in nonlinear conducting media. In this work, we propose a new model for circularly polarized intense EM waves coupled to low-frequency electron density perturbations with higher-order corrections to the nonlocal nonlinearity associated with the EM-driven ponderomotive force in degenerate magnetoplasmas. The temporal chaos reported here could serve as a signature of the development of spatiotemporal chaos in degenerate plasmas.     }
	\section{Introduction} \label{sec-intro}
	The generation of localized wave packets such as electromagnetic (EM) envelope solitons is possible when high-frequency EM waves undergo nonlinear self-modulation, which is generally explained by a nonlinear Schr\"odinger (NLS) Equation  \cite{Zakharov1972Exact, Taniuti1969Modulational}. Relativistic inertia and degeneracy pressure can influence the formation and evolution of such solitons in relativistic degenerate plasmas  \cite{Tidman1996, Markovich1998, ElLabany2020,ElLabany2020,RoyMisra2022}. Both relativistic velocity effects \cite{RoyMisra2022} and relativistic degeneracy pressure \cite{Dey2024} can influence the propagation characteristics of electromagnetic (EM) solitons in nonlinear dispersive media.
	Such EM solitons transport energy and interact nonlinearly with low-frequency electrostatic (ES) perturbations. The resulting dynamics are associated with the energy localization mechanism facilitated by modulational instability (MI), which marks the threshold for soliton formation.
	When the growth rate of MI reaches its critical value, energy localization becomes maximized, and envelope solitons emerge \cite{MisraBhowmik2009}. As the instability saturates or diminishes beyond this point, the energy redistributes, leading to either stabilization or decay of the solitonic structure. In particular, energy exchange between EM solitons and ES perturbations is crucial, as the soliton may lose energy to the ES background or it gets modulated by the ES wave emission. Such interactions can result in a complete energy transfer in conservative systems, transforming coherent soliton structures into incoherent ones and triggering the start of chaos or turbulence \cite{Banerjee2010}. 
	\par
To investigate the complex interactions between EM solitons and electrostatic perturbations, we employ a multi-step modelling approach. Starting from a set of relativistic fluid equations coupled to Maxwell's equations and a degeneracy pressure law, and using Coulomb's gauge condition, we derive a system of coupled nonlinear partial differential equations that describe the full spatiotemporal dynamics (For details, see, e.g., Refs. \cite{misra2018,Holkundkar2018}). Then we apply a low-dimensional truncation method (Galerkin-type approximation), which expands the field variables into a set of finite normal modes to reduce the system for temporal analysis. In this case, we select three resonant modes that capture the primary energy exchange between the EM envelope and the driven electron plasma waves. The validity of the derived autonomous system is then rigorously tested through the linear stability analysis method. For that, we calculate the Lyapunov exponent to quantify the sensitive dependence on initial conditions and Poincar{\'e} sections and power spectra to distinguish between quasiperiodic and chaotic strange attractors. This methodology ensures that the observed transitions are mathematically robust and physically grounded within the subsonic regime of the plasma response. 
	Thus, we consider a simplified three-wave temporal model for the interactions between intense EM waves and the slow response of electrostatic plasma density perturbations. The primary motivation for this system is the nonlinear coupling between a circularly polarized relativistic EM envelope soliton and two ES electron density perturbations, assuming that all other background modes remain unchanged. We construct the model under the constraint of a finite interaction domain, where the wave number $k$ lies within a subinterval that limits the number of interacting modes. One can determine the number of potential excited modes by the spatial domain size, which is inversely proportional to $k$ \cite{Banerjee2010}. This domain guarantees the truncation to a low-dimensional system for three-wave interactions.
	Even with this simplification, the reduced system can exhibit a range of nonlinear phenomena, including temporal chaos \cite{Misra2024, Chow1992}, soliton decay \cite{Kaup1978, Chian1996}, and modulational instability \cite{MisraBhowmik2009}. The chaotic dynamics observed in this regime are not merely artefacts of low-dimensional modeling but can act as precursors to spatiotemporal chaos (STC), which occurs in more extended systems. As shown in previous studies, once a few interacting modes destabilize, they can seed large-scale turbulent structures through energy redistribution \cite{Banerjee2010}.
	\par
	Specifically, we examine how the relativistic degeneracy parameter $R_0\equiv \left(n_0/n_{\rm {cr}}\right)^{1/3}$, where $n_{\rm {cr}}\approx6\times10^{29}~\rm{cm}^{-3}$ denotes the critical value of the particle number density, $n_0$ above which the degeneracy effect starts playing a role,  affects the nonlinear coupling strength by regulating the degree of degeneracy in the electron population. For example, when $R_0=1.2$ (The case of moderate relativistic degeneracy), quasiperiodic-like behavior predominates for $0.35 \lesssim k\lesssim1$, whereas chaotic dynamics are most noticeable in the sub-interval $0 <k \lesssim0.35$. Here, $k$ is the wave number of modulation. A similar bifurcation structure appears for $R_0=0.8$ (The case of weak relativistic degeneracy), but in the sub-intervals, $0.22 \lesssim k \lesssim 1$  and  $0< k \lesssim 0.22$. The Lyapunov exponent spectra and phase space portraits of the interacting wave amplitudes characterize these transitions.
	This investigation provides theoretical insight into how nonlinear EM-ES coupling, under realistic plasma conditions, can result in wave energy transformation, thereby offering a pathway toward understanding wave turbulence in astrophysical and laboratory plasmas. While our present focus is limited to temporal dynamics, the signatures uncovered here may inform future studies of spatiotemporal chaos and turbulence in higher-dimensional systems, including those relevant to laser-plasma interactions \cite{Rososhek2025, Zulick2025} and compact astrophysical objects \cite{Yuan2025, Beattie2025}.
	\par 
Previous studies rigorously examined linearly or elliptically polarized EM waves in nondegenerate plasmas. However, our work addresses the more complex interaction of circularly polarized waves with fully relativistic degenerate electrons. The main mathematical challenge arises from the self-consistent coupling between the EM-driven ponderomotive force and the relativistic degenerate pressure, which yields a more rigorous derivation of the governing NLS equation. Furthermore, we introduce a higher-order correction to the nonlocal nonlinearity. This higher-order term drastically modifies the energy localization threshold and the subsequent stability of EM envelope solitons, a feature that is absent in traditional local (cubic) nonlinear models. We include these effects, which allow us to identify the critical role of the degeneracy parameter $R_0$ in suppressing instabilities and favouring stable soliton evolution in high-density astrophysical environments.  
\par The novelty of this research lies in the formulation of a new coupled model that accounts for circularly polarized, intense EM waves interacting with low-frequency electron density perturbations, while specifically including higher-order corrections to the nonlocal nonlinearity. In the existing literature, which typically assumes a local cubic nonlinearity or focuses on nondegenerate plasma species, our method captures the intricate interplay between relativistic degeneracy and the circularly polarized EM-driven ponderomotive force. This study has significant research importance, as it offers a pathway toward understanding chaos and wave turbulence, as well as energy localization in superdense astrophysical environments, including the interiors of white dwarfs and neutron stars, and in next-generation high-intensity laser-plasma interaction experiments. Furthermore, the investigation of the chaotic dynamics in such systems is essential, as the emergence of temporal chaos in our low-dimensional model provides a path to spatiotemporal chaos and complex wave-wave interactions in complete nonlinear systems. 
	\section{The model and modulational instability} \label{sec-model}
	We consider the nonlinear interaction of circularly polarized intense EM waves with electrostatic electron density perturbations in an unmagnetized plasma  with a flow of relativistic fully degenerate electrons and stationary ions, and assume that all the field variables depend on the space and time coordinates $(z,t)$.   
	The evolution of circularly polarized EM waves in a relativistic plasma with degenerate electrons is governed by the following equation \cite{Holkundkar2018}.  
	
	\begin{equation} \label{eq1}
		\Bigg(\frac{\partial^2}{\partial t^2}-c^2\frac{\partial^2}{\partial z^2}\Bigg)a= -\frac{\omega_p^2}{n_0}\frac{n}{\gamma}a,
	\end{equation}
	where $c$ is the speed of light in vacuum, $n$ is perturbed electron number density in laboratory frame with $n_0$ denoting its unperturbed value, and $\omega_p=\sqrt{n_0e^2/\varepsilon_0m}$ is the electron plasma oscillation frequency with $e$ denoting the elementary charge, $m$ the electron mass, and $\varepsilon_0$ the permittivity of free space. Also,   $a = eA/mc^2$ is the dimensionless EM wave amplitude and $\gamma$ is the relativistic Lorentz factor, which for slow motion approximation of relativistic degenerate electrons in the EM wave fields (but full relativistic effects on the thermal motion) reads \cite{misra2018}
	\begin{equation}
		\gamma = \sqrt{\frac{1+e^2n_0^2A^2/H_0^2}{1-v^2/c^2}}\approx 1+\frac{1}{2}\left(\frac{en_0A}{H_0}\right)^2.
	\end{equation}
	Here, $A$ is the wave vector potential, $v$ is the longitudinal electron fluid velocity along the $z$-axis, and $H_0=n_0mc^2\sqrt{1+R_0^2}$ is the enthalpy per unit volume of the fluid with $R_0\equiv p_{\rm{Fe}}/mc=\left(n_0/n_{\rm{cr}}\right)^{1/3}$ denoting the relativistic degeneracy parameter defined in Sec. \ref{sec-intro}. The regime with $R_0\ll1$ corresponds to weakly relativistic degenerate plasma, whereas ultra-relativistic degenerate plasmas are characterized by $R_0\gg1$. The values in between these limits can be considered as moderate or strong degenerate plasmas.
\par 
Equation \eqref{eq1} can be viewed as the fundamental wave equation for the electromagnetic (EM) vector potential in a relativistic plasma. This equation describes how a high-intensity laser pulse or an EM signal evolves as it interacts with the electron-density perturbations. The right-hand side of Eq. \eqref{eq1}  represents the coupling between the EM field and the plasma waves, where $n/\gamma$ captures the relativistic and density-dependent modifications. In the context of laser wakefield generation, this equation is vital for predicting the pulse and the resulting formation of soliton structures \cite{Holkundkar2018}. Furthermore, in astrophysical contexts \cite{Misra2024}, it models the dynamics of intense radiation pulses emitted by superdense objects, such as white dwarfs, where the plasma's relativistic motion and degeneracy significantly alter wave propagation. 	
	\par    
	For slowly varying weakly nonlinear circularly polarized EM wave envelopes (i.e., $|{\partial a}/{\partial t}| \gg |{\partial^2 a}/{\partial t^2}|, |a|\omega^2$, where $\omega$ is the EM wave frequency), assuming the wave vector potential to be of the form,
	\begin{equation} \label{eq-a}
		{\vec{a}} = \frac{1}{2} a(z,t) e^{i\omega t} (\hat{x}-i\hat{y}) + \mathrm{c.c.},  
	\end{equation} 
	where ``c.c." denotes the complex conjugate, we obtain from Eq. \eqref{eq1}, the following nonlinear Schr{\"o}dinger (NLS) equation with nonlocal nonlinearity for slowly varying EM envelope solitons. 
		\begin{equation}
		i\frac{\partial a}{\partial t} + \frac{1}{2}\frac{c^2}{\omega} \frac{\partial^2 a}{\partial z^2}  = \frac{1}{2} na \bigg(\frac{\omega_p^2 }{\omega {n_0}}\bigg) (1- \sigma a^2), \label{eq2}
	\end{equation}
	where $\sigma=1/2\left(1+R_0^2\right)$ corresponds to a small correction to the nonlcal nonlinearity. We note that the higher the values of $R_0$, the smaller are the $\sigma$-values, implying that the nonlocal nonlinear response tends to become higher in strong or ultra-relativistic degenerate regimes. While the cubic nonlinearity tends to favor the formation of envelope solitons, the nonlocal nonlinearity can significantly modify the soliton's stability and form.
	\par  Equation \eqref{eq2} is to be coupled with the following equation for the low-frequency electron density perturbations that are driven by the EM wave's ponderomotive force  \cite{misra2018}. 
	\begin{equation}
		\Bigg(\frac{\partial^2}{\partial t^2}-\delta\frac{\partial^2}{\partial z^2}+1\Bigg) n = \frac{1}{2} (1-\delta) \frac{\partial^2 a^2}{\partial z^2}, \label{eq3}
	\end{equation}
	where $\delta=\left(1-2\sigma/3\right)$ depends on the degeneracy parameter $R_0$ and contributes to the wave dispersion.
	\par 
	After normalizing the variables according to, $ t \rightarrow t\omega, ~z \rightarrow z\omega /c$, and  $n \rightarrow n/n_0$, we recast Eqs. \eqref{eq2} and \eqref{eq3} in the following dimensionless form.
	\begin{equation}
		i\frac{\partial a}{\partial t}+ \frac{1}{2}\frac{\partial^2 a}{\partial z^2} = \frac{1}{2} \beta_0^2 na\left(1-\sigma |a|^2\right), \label{eq4}		
	\end{equation}
	\begin{equation}
		\Bigg(\frac{\partial^2}{\partial t^2}-\delta\frac{\partial^2}{\partial z^2}+\eta\beta_0^2 \Bigg) n = \frac{1}{2} \eta^2 \left(1-\delta\right) \frac{\partial^2 |a|^2}{\partial z^2} ,	\label{eq5}
	\end{equation}
	where $\eta = \sqrt{2\sigma}$ and $\beta_0 = \omega_p/\omega$ is the ratio between the frequencies of constant plasma oscillation and the EM wave. Equations \eqref{eq4} and \eqref{eq5} are the desired set of equations for the nonlinear interaction between EM wave envelopes and electron plasma density perturbations in an unmagnetized plasma with relativistic flow of degenerate electrons.
The novelty of the model equations \eqref{eq4} and \eqref{eq5} lies in the specific formulation of the coupling and the dispersion terms, which incorporates the relativistic degeneracy. Equation \eqref{eq4} introduces a higher-order nonlocal nonlinear correction $\sigma|a|^2$, where $\sigma$  uniquely introduces the degeneracy parameter $R_0$. Unlike the standard cubic NLS equation used in previous studies of nondegenerate plasmas, our model accounts for the shifting balance of self-focusing effects in superdense environments. Furthermore, the driven density perturbation equation \eqref{eq5} includes the dispersion coefficient $\delta$, which couples the electrostatic response directly to the electron's relativistic enthalpy. Now the derived coupled system [Eqs. \eqref{eq4} and \eqref{eq5}] provides the first theoretical framework to investigate how relativistic motion and relativistic degeneracy of electrons specifically influence the transition to chaos in circularly polarized EM wave envelopes. 	
\par 
For the validity of Eqs. \eqref{eq4} and \eqref{eq5}, we mention that Eq. \eqref{eq4} agrees with Eq. (13) of Ref. \cite{Holkundkar2018} when one considers a linearly polarized EM wave [instead of a circularly polarized wave, see Eq. \eqref{eq-a}], nondegenerate electrons (instead of degenerate electrons), and ignores the higher-order nonlinearity (proportional to $\sigma$). Also, Eq.  \eqref{eq5} agrees with Eq. (21) of Ref. \cite{misra2018} after a minor adjustment with the parameter and the dependent variables. Thus, Eqs. \eqref{eq4} and \eqref{eq5} are consistent with previous investigations on the generation of electromagnetic solitons \cite{Holkundkar2018,misra2018,roy2020}. Furthermore, authors in Ref. \cite{Holkundkar2018} have validated the relativistic fluid framework employed here with Particle-In-Cell (PIC) simulations,  demonstrating excellent agreement in describing the transition from wakefield generation to soliton formation. The physical reliability of our model gets further strengthened by the modulational instability (MI). We have observed that the onset of chaotic states becomes intrinsically linked to the maximum growth rate of the MI [Eq. \eqref{eq-MI}], ensuring that the temporal dynamics are a direct consequence of the physical instability threshold rather than numerical artifacts. 
\par
\begin{figure*}[!ht]
		\centering
		\includegraphics[width=6.5in,height=2.5in]{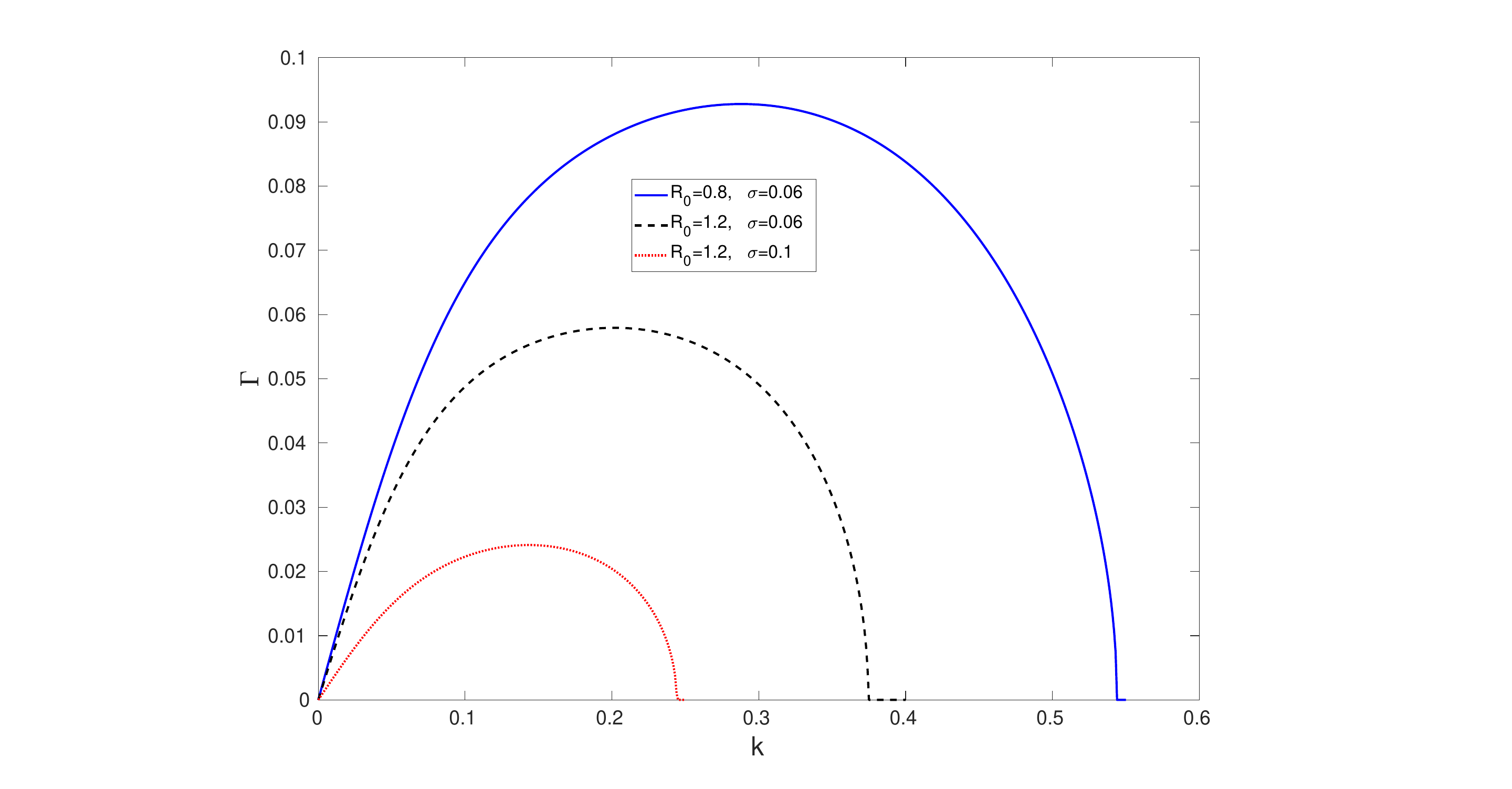}
		\caption{The modulationsl instability growth rate $\Gamma$ is plotted against the modulation wave number ($k$) for different values of the degeneracy parameter $R_0$ and $\sigma$ as in the legend.}
		\label{fig:growth-label}    
	\end{figure*}

 
Equations \eqref{eq4} and \eqref{eq5} can have potential applications in studying several nonlinear phenomena, including the evolution of EM envelope solitons through the modulational instability \cite{RoyMisra2022}, generation of wakefields \cite{Holkundkar2018,roy2020}, as well as the emergence of temporal \cite{misra2010} and spatiotemporal chaos \cite{Banerjee2010} that can lead to the onset of turbulence in nonlinear dispersive media \cite{Zakharov1972Exact}. However, we have limited our study to the emergence of chaos in a low-dimensional temporal model in regimes of the modulation wave number where the instability growth rate increases, reaches a maximum, and then vanishes due to the effects of the degeneracy parameter $R_0$ and the nonlinear correction parameter $\sigma$. Since modulational instability is a well-known mechanism for the generation of envelope solitons, before proceeding to study the temporal dynamics of EM solitons, it is pertinent to examine the condition for MI and the instability growth rate as a function of modulation wave number.    
	So, looking for the modulation of EM wave amplitude and making the ansatz, 
	\begin{equation} \label{eq-ansatz}
		\begin{split}
			&a(z,t)=\bar{a}(z,t)\exp[i\theta (z,t)],\\
			&n(z,t)=\tilde{n}(z,t)\exp\left(ikz-i\omega t\right)+\mathrm{c.c.},\\
			&\bar{a}(z,t)=a_0+\tilde{a}\exp\left(ikz-i\omega t\right)+\mathrm{c.c.},\\
			&\theta(z,t)=\theta_0+\tilde{\theta}\exp\left(ikz-i\omega t\right)+\mathrm{c.c.}, 
		\end{split}
	\end{equation}
	where $a$ and $\theta$ are slowly varying functions of $z$  and $t$, and $\tilde{a}\ll a_0$, $\tilde{\theta}\ll \theta_0$, we obtain from Eqs. \eqref{eq4} and \eqref{eq5} the  following linear dispersion relation.
	\begin{equation}
		\begin{split}
			&(\omega^2-\delta k^2 -\eta \beta_0^2)(\omega^2-\frac{1}{4} k^4)\\
			&-\frac{1}{4} a_0^2 \beta_0^2 \eta^2 k^2 (1-\delta)(1-\sigma a_0^2)= 0. \label{eq-disp}
		\end{split}
	\end{equation}
	From Eq. \eqref{eq-disp}, we find that the first two factors on the first term in the left side, respectively, correspond to the dispersion modes for electron plasma oscillations and the EM wave, and they get coupled due to the EM wave ponderomotive force proportional to $a_0^2$ (See the second term on the left side). In the absence of the latter, the modes get decoupled and there will be no instability and hence no formation of a EM envelope soliton. Thus,
	looking for the MI, and assuming $\omega \simeq i\Gamma$, we obtain the following expression for the instability growth rate.
	\begin{equation} \label{eq-MI}
		\Gamma = \frac{1}{2} k \Bigg[ \frac{a_0^2 \zeta_0^2}{k^2\delta -{ k^4}/{4}+\eta \beta_0^2} -k^2\Bigg]^{1/2},
	\end{equation}
	where $\zeta_0 = \beta_0^2 \eta^2 (1-\delta)(1-\sigma a_0^2)$, and the expression under the square brackets is positive for some $k<k_c$ (at which the MI sets in) with $k_c$ denoting the critical value of $k$.  The growth rate attains a maximum at $k\approx k_c/\sqrt{2}$. Figure \ref{fig:growth-label} shows the profiles of MI growth rate against the wave number of modulation for different values of the degenerate parameter $R_0$ and the nonlocal correction parameter $\sigma$ as in the legends. The growth rate initially increases with $k$, reaches a maximum value, and then decreases to a cutoff at $k=k_c$. An increase in $R_0$ causes a transition from the weak $(R_0<1)$ to the strong $(R_0>1)$ degenerate regime. A slight increase in $\sigma$ reduces the nonlocal nonlinearity. Any one of these changes can significantly decrease the growth rate and cause a cutoff at a lower wave number (See the dashed and dotted lines). Therefore, higher values of $R_0$ and $\sigma$ result in a lower growth rate and shift $k_c$ to smaller values. Thus, we may conclude that relativistic degenerate plasmas with higher degeneracy (i.e., in the density regimes beyond the critical density, $n_{\rm{cr}}\approx 6\times10^{29}~\rm{cm}^{-3}$) favor the formation of more stable and well-defined EM solitons. If the EM wave field is so strong that the MI threshold exceeds that of the decay instability, the EM waves get trapped by the electron density perturbations. The nonlinear interaction between these waves may result in turbulence, during which wave energy is redistributed or transferred from lower to higher harmonic modes \cite{Banerjee2010}. However, since we have limited our study to the temporal dynamics of EM waves, one can observe these features in the full-dimensional model [Eqs. \eqref{eq4} and \eqref{eq5}, but beyond the scope of the present work. 
	\par 
	Equations \eqref{eq4} and \eqref{eq5} are, in general, multidimensional. They feature the excitation of an infinite number of EM wave modes. However, it is reasonable to assume that only a few modes participate actively in the nonlinear wave-wave interactions. A low-dimensional model with a few (namely, three) truncated modes helps study the basic features of the full-wave dynamics of Eqs. \eqref{eq4} and \eqref{eq5}. We must note, though, that the precise details of the low-dimensional model depend strongly on the range of the wave number of modulation $k$. We will note shortly that lowering $k$ beyond some limits in $0<k<k_c/\sqrt{2}$ will result in several modes larger than three, which will question the validity of a three-wave interaction model. Thus, we consider the nonlinear interactions among three wave modes by expanding the EM wave vector potential $a(z,t)$ and the electron density perturbation $n(z,t)$ as
	\begin{equation} \label{eq9}
		\begin{split}
			a(z,t)=&\sum_{m=-M/2}^{+M/2}a_m(t)e^{imkz},\\
			=& \sum_{m=-M/2}^{+M/2}\rho_m(t)e^{\theta_m(t)}e^{imkz},\\
			=&a_0(t)+a_{-1}(t)  e^{-ik z}+a_{1}(t)  e^{ik z},		
		\end{split}
	\end{equation} 
	\begin{equation}
		\begin{split}
			n(z,t)=&\sum_{m=-M/2}^{+M/2}n_m(t)e^{imkz},\\
			=&n_0+n_{1}(t)  e^{ik z}+n_{-1}(t)  e^{-ik z},		\label{eq10}
		\end{split}
	\end{equation}
	where $M=[k^{-1}]$ denotes the number of modes to be selected in the interactions. Also,   $a_{-m}=a_{m}$ and $n_{-m}=n_{m}$. In the case of three-wave interactions, we choose $M=2$. In a sense, Eqs. \eqref{eq9} and \eqref{eq10} describe the general solution of Eqs. \eqref{eq4} and \eqref{eq5}  as a  superposition  of a set of finite normal modes. 
			\par
	Substituting from Eqs. \eqref{eq9} and \eqref{eq10} the expressions for $a$ and $n$ into Eqs. \eqref{eq4} and \eqref{eq5}, and following the same approach as Refs. \cite{misrapla2008,Sharma2005}, we obtain 
	\begin{equation}
		\begin{split}
			i\dot{a}_0 = \frac{1}{2}\beta_0^2 \{p_1- (p_1p_1^{\prime}+p_2p_3^{\prime}+p_3p_2^{\prime}+p_4p_4^{\prime}+p_5p_5^{\prime})\},
		\end{split}				\label{eq-b5}	
	\end{equation} 
	\begin{equation}
		i\dot{a}_1 = \frac{1}{2}\beta_0^2 \{p_2-(p_1p_2^{\prime}+p_2p_1^{\prime}+p_3p_5^{\prime}+p_4p_3^{\prime}) \}, \label{eq-b6}
	\end{equation}
	\begin{equation}
		i\dot{a}_{-1} = \frac{1}{2}\beta_0^2 \{p_3-(p_1p_3^{\prime}+p_3p_1{\prime}+p_5p_2^{\prime}+p_2p_4^{\prime}) \}, \label{eq-b7}
	\end{equation}
	where $p_1=a_1n_0+a_{-1}n_1+a_1n_1^{*}$, $p_2=a_1n_0+a_0n_1$, $p_3=a_1n_0+a_0n_1$, $p_4=a_1n_1$, $p_5=a_{-1}n_1^{*}$, $p_1^{\prime}=\sigma(a_0a_0^{*}+a_1a_1^{*}+a_1a_1^{*})$, $p_2^{\prime}= \sigma(a_0a_1^{*}+a_1a_0^{*})$, $p_3^{\prime}=\sigma(a_0a_1^{*}+a_{-1}a_0^{*})$, $p_4^{\prime}=\sigma a_{-1}a_1^{*}$, and $p_5^{\prime}=\sigma a_1a_{-1}^{*}$. Also, the dot denotes differentiation with respect to $t$ and the asterisk denotes the complex conjugate.
	\par 
	We multiply Eq. \eqref{eq-b5} by $a_0^*$ to form a new equation and subtract it from its complex conjugate to get
	\begin{equation}\label{eq-b10}
		\begin{split}
			i \left| \dot{a_0} \right|^2 = & 2\sigma (a_0 a_1^* - a_1 a_0^{*}) (n_0 + n_1 a_1 a_1^*)- \\
			&\left\{ 1 - \sigma (a_0 a_0^* + 2 a_1 a_1^*) \right\} (n_0 + 2n_1)(a_0 a_1^{*} - a_1 a_0^{*}).
		\end{split}
	\end{equation}
	Similarly, applying the same operation for  Eqs. \eqref{eq-b6} and \eqref{eq-b7}, and adding all the three resulting equations like Eq. \eqref{eq-b10}, we get 
	\begin{equation}
		|a_{-1}|^2+|a_{1}|^2+|a_0|^2=N,												\label{eq-b11}
	\end{equation}
	where  we assume  $n_1=n_1^\ast,~ n_0=N$  (the plasmon number), and introduce  the new variables, $\rho_{0},~ \rho_{1},~ \theta_{0}$, and $\theta_{1}$ according to $a_0=\rho_{0}e^{i\theta_{0}}$,  $a_{-1}=a_0=\rho_{1}e^{i\theta_{1}}$, $\rho_0=\sqrt{n_0}\sin{w}$, $\rho_1=\sqrt{n_0}\cos{w}$, $\psi=2w$.
	Substituting the expressions \eqref{eq9} and \eqref{eq10} for $a$ and $n$ into Eq. \eqref{eq5}, and using the new  variables as defined above, we get 
	\begin{equation} \label{ss-1}
		\begin{split}
			\ddot{n_1}&=-(\delta k^2+\beta_0^2\eta)n_1+\frac{1}{2}\eta^2(1-\delta)n_0\sin{\psi}\cos{\phi}.	
		\end{split}
	\end{equation} 
	Also, from Eqs. \eqref{eq-b10} and from Eqs. \eqref{eq-b6} and \eqref{eq-b7} using the newly defined variables,  we successively obtain 
	\begin{equation} \label{ss-2}
		\begin{split}
			\dot{\psi} = \beta_0^2 \sin{\phi} [\{(2+N)\sigma-1\}N+2(\sigma N-1)n_1& \\ +\sigma N(N+4n_1)\cos^2{(\psi/2)}],
		\end{split}
	\end{equation}			
	\begin{equation} \label{ss-3}
		\begin{split}
			\dot{\phi} = \frac{1}{2}(k^2+\sigma N^2 \beta_0^2)
			+\frac{1}{2} \beta_0^{2} n_1(\cos{\frac{\psi}{2}} \cos{\phi}-1)&\\+ \frac{1}{2} N\sigma \beta_0^2 \cos{\phi} \tan{\frac{\psi}{2}},
		\end{split}
	\end{equation}
	where $\phi=\theta_0-\theta_1$. 
	\par 
	 Equations \eqref{ss-1}-\eqref{ss-3} constitute the required low-dimensional model for three-wave interactions of circularly polarized high-frequency EM waves and low-frequency electron plasma oscillations driven by the EM wave ponderomotive force. We note that these equations are significantly modified by electron degeneracy pressure through the parameter $\eta$ and by the nonlocal correction proportional to $\sigma$. Although these equations look similar to Ref. \citep{sroy2023chaos}, the relevant dynamics and evolution of envelope solitons are distinctive. For the sake of convenience, we recast, after redefining the variables, $\psi = x$, $n_1= y$, $\dot{y}=z$, and $\phi=w$ as
		\begin{align} \label{eq-reduced}
		\dot{x} &= -\beta_0^2 \sin w 
		\Big[ \{(2+N)\sigma - 1\}N 
		+ 2(\sigma N - 1)y  \notag \\
		&\quad + \sigma N(N + 4y) \cos^2\!\left(\frac{x}{2}\right) \Big], \notag \\
		\dot{y} &= z, \notag \\
		\dot{z} &= -(\delta k^2 + \beta_0^2 \eta)y
		+ \frac12 \eta^2 (1 - \delta) N \sin x \cos w, \notag \\
		\dot{w} &= \frac12 \big(k^2 + \sigma N^2 \beta_0^2 \big)
		+ \frac12 \beta_0^{2} y \left( \cos\frac{x}{2} \cos w - 1 \right) \notag \\
		&\quad + \frac12 N\sigma \beta_0^2 \cos w \tan\frac{x}{2}.
	\end{align}
	Before we proceed to study the nonlinear dynamical features of the system of equations \eqref{eq-reduced} in Sec. \ref{sec-nonlinear}, it is pertinent to study the linear stability of the system about the equilibrium points (See Sec. \ref{sec-linear}).     
	\section{Linear stability analysis} \label{sec-linear}
		In order to perform the stability analysis of the system \eqref{eq-reduced}, we first find its equilibrium points. To this end, we equate the right-hand sides of Eq. \eqref{eq-reduced} to zero and find solutions for $x,~y,~z,~w$ as $(x_{10}, y_{10}, z_{10}, w_{10})$. One  such point is $P_1\equiv(0,0,0,\frac{1}{2}(k^2+\sigma N^2 \beta_0^2))$.  Some other equilibrium points can obtained by solving the following two sets of equations. 
	\begin{equation} \label{fixed-1}
		\begin{split}
			&(k^2+\sigma N^2 \beta_0) =\beta_0^2[(-1)^n(y\cos{\frac{x}{2}}+\sigma N\tan{\frac{x}{2}})-y], \\
			&2(\delta_e k^2+\beta_0^2 \eta_e) = (-1)^n N\{\eta_e^2(1-\delta_e)\}\sin{x},\\
			&\omega =n\pi, ~z=0,
		\end{split}
	\end{equation}
		\begin{equation} \label{fixed-2}
		\begin{aligned}
			&\{(2+N)\sigma - 1\}N 
			+ 2(\sigma N - 1)y  \\
			&\quad + \sigma N(N + 4y) \cos^2\!\left( \frac{x}{2} \right) = 0, \\
			&(\delta_e k^2 + \beta_0^2 \eta_e) y
			= \frac12 \eta_e^2 N (1 - \delta_e) \sin x \cos w, \\
			&(k^2 + \sigma N^2 \beta_0^2) 
			+ \beta_0^2 y \left( \cos\frac{x}{2} \cos w - 1 \right) \\
			&\quad + N\sigma\beta_0^2 \cos w \tan\frac{x}{2} = 0.
		\end{aligned}
	\end{equation}
	%
		However, we will perform the stability analysis around $P_1$. To this end, we apply the transformation: $x'=x-x_{10}$, $y'=y-y_{10}$, $z'=z-z_{10}$ and $w'=w-w_{10}$ to obtain a linearized system of perturbation equations
	\begin{equation}
		\frac{dX}{dt}=JX,
	\end{equation}
	where $J$ is the jacobian matrix and $X=(x',y',z',w')$. We obtain the eigenvalues $(\lambda)$ from the equation $JX=\lambda X$ and study the stability of the system \eqref{eq-reduced} by the nature of the eigenvalues.
	The Jacobian matrix corresponding to the equilibrium point $P_1=(0,0,0,\frac{1}{2}(k^2+\sigma N^2 \beta_0^2))$ is given by
	\begin{widetext}
		\begin{equation}
			J|_{P_{1}} =
			\begin{bmatrix}
				0 & \beta_0^2 \sin{w^*} (6\sigma N - 2) & 0 & \beta_0^2 \cos{w^*} \left[ ((2+N)\sigma - 1)N + \sigma N^2 \right] \\
				0 & 0 & 1 & 0 \\
				\frac{1}{2} \eta_e^2 (1 - \delta_e) N \cos{w^*} & -(\delta_e k^2 + \beta_0^2 \eta_e) & 0 & 0 \\
				\frac{1}{4} N \sigma \beta_0^2 \cos{w^*} & \frac{1}{2} \beta_0^2 (\cos{w^*} - 1) & 0 & 0 \\
			\end{bmatrix},
		\end{equation}
	\end{widetext}
	where $w^{*}= \left(k^2+\sigma N^2 \beta_0^2\right)/2$.
	The characteristic equation corresponding to the matrix $J|_{P_{1}}$ is
	\begin{equation} \label{ch-eq}
		16\lambda^{4}-4(D+BE)\lambda^2-2AC\lambda-B(CF-DE) =0, 
	\end{equation}
	where the coefficients are 
	\begin{equation}
		\begin{aligned}
			A &= \beta_0^2 \sin{\left( \frac{1}{2}(k^2 + \sigma N^2 \beta_0^2) \right)}  (6\sigma N - 2), \\
			B &= \beta_0^2 \cos{\left( \frac{1}{2}(k^2 + \sigma N^2 \beta_0^2) \right)}  \left[ ((2+N)\sigma - 1)N + \sigma N^2 \right], \\
			C &= \frac{1}{2} \eta_e^2 (1 - \delta_e) N \cos{\left( \frac{1}{2}(k^2 + \sigma N^2 \beta_0^2) \right)}, \\
			D &= -\left( \delta_e k^2 + \beta_0^2 \eta_e \right), \\
			E &= \frac{1}{4} N \sigma \beta_0^2 \cos{\left( \frac{1}{2}(k^2 + \sigma N^2 \beta_0^2) \right)}, \\
			F &= \frac{1}{2} \beta_0^2 \left( \cos{\left( \frac{1}{2}(k^2 + \sigma N^2 \beta_0^2) \right)} - 1 \right).
		\end{aligned}
	\end{equation}
	\par
\begin{figure*}[!ht]
		\centering
		\includegraphics[width=0.6\textwidth]{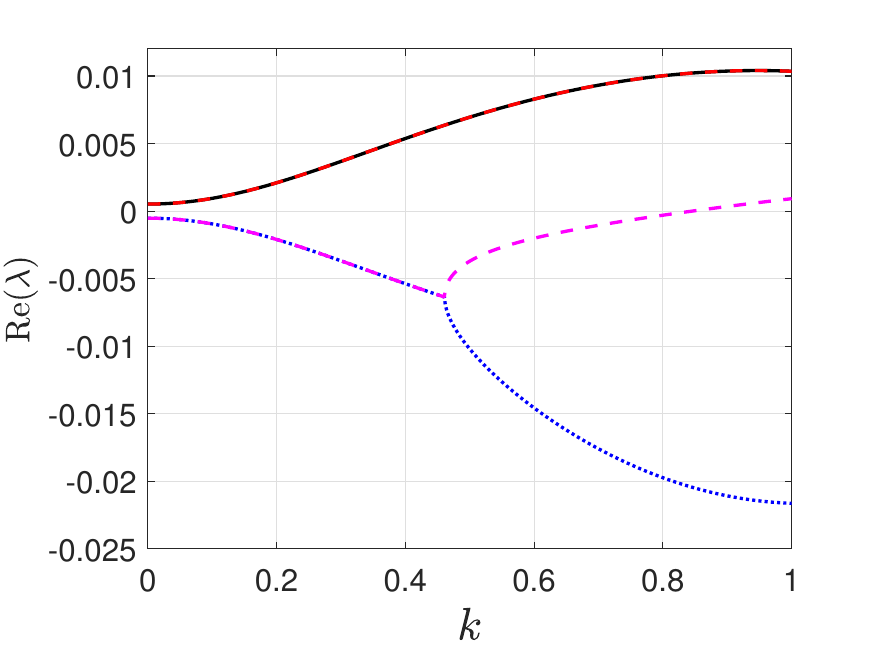}
		\caption{The real parts of the eigenvalues corresponding to Eq. \eqref{ch-eq} are shown for the parameters, $R_0=1.2,~\sigma=0.16,~\beta_0=0.34$, and $N=1$. We observe that there is at least one positive eigenvalue in the entire domain of $k$, predicting the existence of chaos in the temporal model [Eq. \eqref{eq-reduced}]. The overlapping of two curves with $\Re\lambda>0$ appears due to a pair of complex-conjugate eigenvalues.   }
		\label{fig-eigen}
	\end{figure*}

	We numerically study the nature of the real parts of the roots of Eq. \eqref{ch-eq} corresponding to the equilibrium point $P_1$. Figure \ref{fig-eigen} displays the profiles of the real parts of four eigenvalues: Two of them are positive and identical (See the solid red and dashed black lines) in the entire domain of $k$ and the other two (while one of them is zero or negative, the other can assume zero, negative, or positive values) bifurcate at a point within $0.4<k<0.5$. Thus, depending on the values of $k$, the eigenvalues can assume zero, negative, or positive values, implying that the system can evolve into stable (with $\Re\lambda=0$) or unstable (with $\Re\lambda>0$) states about the equilibrium point $P_1$. Since the MI of the EM wave envelopes occurs in $0<k<k_c$, such values of $k$ would provide an initial guess for the domains of the existence of periodic, quasiperiodic, or chaotic states in the wave-wave interactions.  
	\section{Lyapunov exponents, bifurcation diagrams, and phase-space portraits} \label{sec-nonlinear}
	 In Sec. \ref{sec-linear}, we have predicted the regions for the linear stability and instability of the dynamical system \eqref{eq-reduced} in the domains of the wave number of modulation $k$. Next, we proceed to find the ranges of values of the parameters $k$, $\sigma$, and $R_0$  in which the periodic, quasiperiodic, or chaotic states of EM waves can exist by the analysis of Lyapunov exponent spectra $\Lambda_i$, $i=1,2,3,4$, and bifurcation diagrams in the parameter space. We present here some underlying basic theory of the Lyapunov exponent for the sake of clarity \cite{sroy2023chaos}. To find the Lyapunov exponents, we recast the dynamical system \eqref{eq-reduced} in the form $\dot{x}_i=f_i(X)$ with the initial condition,  $X(0)=\left[x(0),~y(0),~z(0),~w(0)\right]$. The existence of different attractors, namely periodic, quasiperiodic, or chaotic phase portraits, typically depends on the initial conditions and how they correspond to different values of the Lyapunov exponents. The latter characterizes the behavior of the state variable $X(t)$ in the tangent space of the phase space, defined by the following Jacobian matrix. 
	\begin{equation}
		J_{ij}(t)=\frac{df_i}{dx_j}\biggr|_{X(t)}. 
	\end{equation} 
	We define the  tangent vectors by the matrix $A$, given by,
	\begin{equation}
		\dot{A}=JA,
	\end{equation}
	with the initial condition $A_{ij}(0)=\delta_{ij}$, where $\delta_{ij}$ denotes the Kronecker delta. Typically, the matrix $A$ characterizes how a small change in the distance between two trajectories in phase space evolves from the initial state $X(0)$ to the final state $X(t)$.  Here, the matrix $A$ is the same as $J|_{P_1}$, obtained before.
	Then we obtain the Lyapunov exponents $\Lambda_i$ as the eigenvalues of the following matrix  \cite{sroy2023chaos}. 
	\begin{equation}
		\Lambda = \lim_{t\rightarrow \infty}\frac{1}{2t}\log\left[A(t) A^T(t)\right], \label{eq-lyap}
	\end{equation}
	where the prefix $T$ denotes transpose of the corresponding matrix. Now, given an initial condition $X(0)$, we can obtain the separation distance between two trajectories in phase space by the Liouville's formula: $\delta X(t)=\mathrm{tr}\left(J(t)\right)|A(t)|$, where $|A(t)|\equiv\mathrm{det}~A(t)$ and $\mathrm{det}~A(0)=1>0$. Thus, for the dynamical system \eqref{eq-reduced}, if we obtain  $\mathrm{det}~A(t)=\exp\left(\int_0^t\mathrm{tr}\left(J(t)\right)dt\right)=1>0$, i.e., at least one $\Lambda_i>0$, there is a possibility for the existence of a chaotic state in a given time interval $[0,t]$.
	\par 
	 Before we analyze the Lyapunov exponent spectra, the bifurcation diagram in parameter space, and the phase-space portraits, we recall a key point. The modulational instability of circularly polarized EM envelopes occurs for $0<k<k_c$. The instability growth rate gradually increases in the interval $0<k<k_c/\sqrt{2}$. It then decreases in $k_c/\sqrt{2}<k<k_c$ until vanishing at $k=k_c$. This implies that the nonlinear wave-wave interaction dynamics are subsonic in the interval $k_c/\sqrt{2}<k<k_c$. However, if $k$ reduces below $k_c/\sqrt{2}$, there may be excitation of many unstable wave modes due to mode selection with $M=[k^{-1}]$. As a result, the relevant dynamics may no longer be subsonic. In this context, a three-wave interaction model remains a useful approximation \cite{sroy2023chaos}. Here, we assume that one EM wave mode ($|m|=1$) is unstable, while the others ($|m|>1$) are stable. Also, two driven electron plasma waves excited by the unstable EM mode also remain. This consideration leads to the autonomous system \eqref{eq-reduced}. The main purpose of this investigation is to study the dynamical features as $k$ decreases from $k_c/\sqrt{2}$ to a certain value (since lower $k$ means higher mode number $M$) and increases from $k_c/\sqrt{2}$ to $k_c$ in the subsonic region where the three-wave model is sustainable. 
	\par
\begin{figure*}[!ht]
		\centering
		\includegraphics[width=0.9\textwidth]{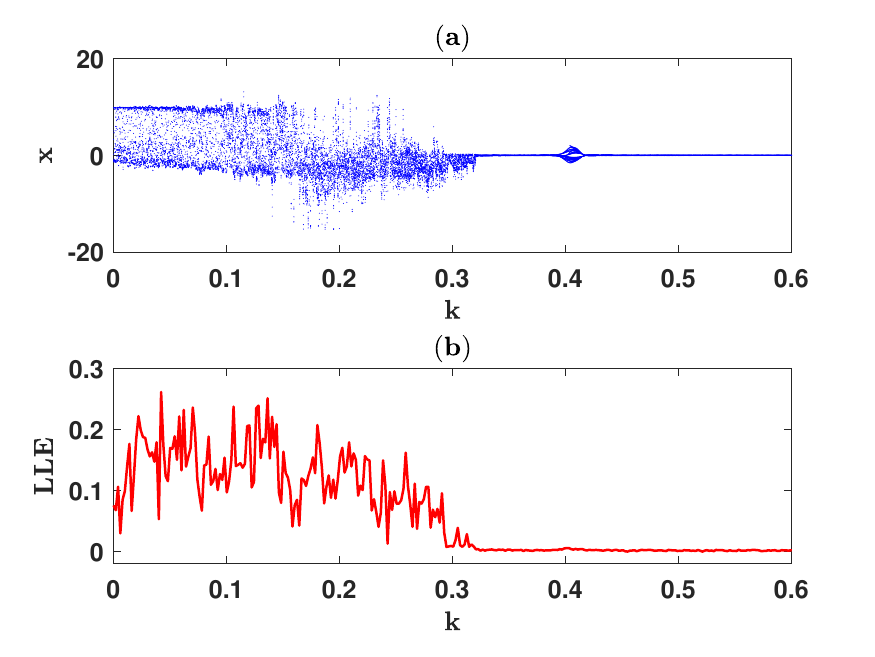}`
		\caption{Bifurcation diagram and the LLE are shown for the set of parameters, $R_0=0.8,~\sigma= 0.06,~N=1$, and $\beta_0=0.05$.}
		\label{fig-bifur-lyap1}
	\end{figure*}

\begin{figure*}[!ht]
		\centering
		\includegraphics[width=0.9\textwidth]{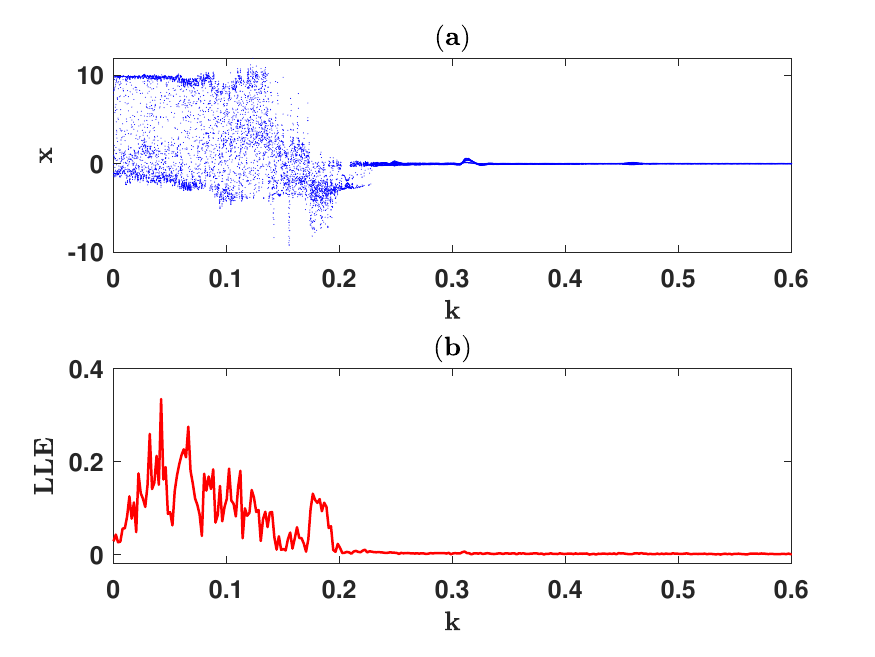}
		\caption{The same as in Fig. \ref{fig-bifur-lyap1} but for a different value of $R_0$, $R_0=1.2$.  }
		\label{fig-bifur-lyap2}
	\end{figure*}

\begin{figure*}[!ht]
		\centering
		\includegraphics[width=0.9\textwidth]{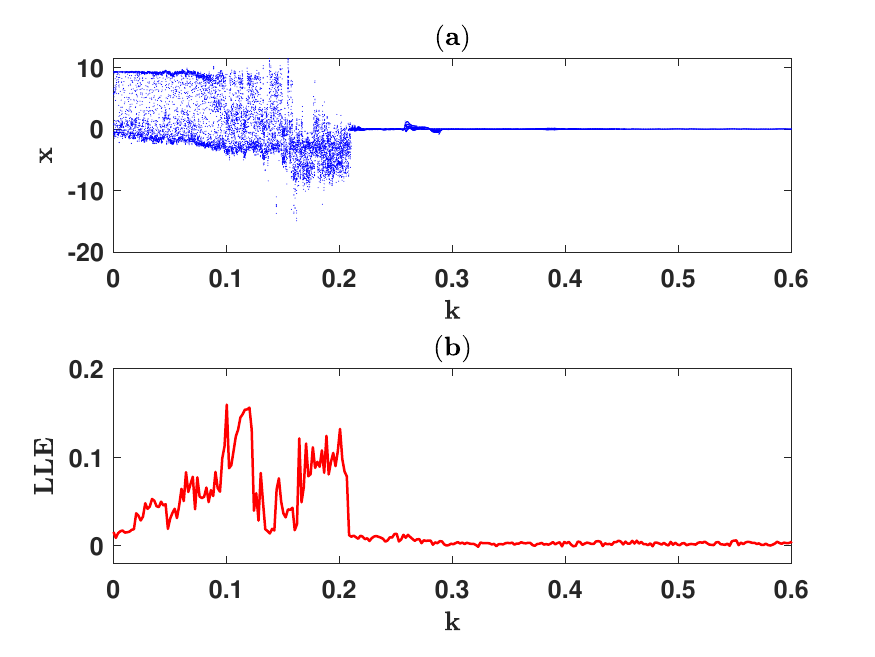}
		\caption{The same as in Fig. \ref{fig-bifur-lyap1} but for a different value of $\sigma$, $\sigma= 0.12$.}
		\label{fig-bifur-lyap3}
	\end{figure*}

 
	 We numerically analyze the bifurcation diagram and calculate the largest Lyapunov exponent (LLE) for Eq. \eqref{eq-reduced} in the domain $0<k<k_c$ with different sets of values for $R_0$ and $\sigma$, and display the results in Figs. \ref{fig-bifur-lyap1}--\ref{fig-bifur-lyap3} where the subplot (a) exhibits the bifurcation diagram (upper panel) and the subplot (b) (lower panel) is for the LLE.  Specifically, there appear to be two domains of $k$ values, in one of which the LLE assumes positive values, and in the other, it is close to zero (but non-negative). While the former confirms the existence of chaos, the latter corresponds to quasiperiodic states in the respective $ k$-domains. These values highly correlate with the corresponding bifurcation diagrams  [Figs. \ref{fig-bifur-lyap1} and \ref{fig-bifur-lyap2}]. We observe that as the degeneracy parameter $R_0$ increases from weakly relativistic $(R_0<1)$ to strong $(R_0>1)$ degenerate regimes (keeping $\sigma$ fixed), the chaotic domain shrinks. Consequently, the existence domain for the quasiperiodic states increases. It follows that the EM wave envelopes under the slow response of the electron plasma perturbations tend to become stable in high-density plasmas with a flow of relativistic, degenerate electrons. We also observe similar phenomena as the nonlocal parameter $\sigma$ increases (keeping $R_0$ fixed), i.e., the contribution from the nonlocal nonlinearity decreases (See Figs. \ref{fig-bifur-lyap1} and \ref{fig-bifur-lyap3}). Typically, the strong nonlocal nonlinearity destabilizes an envelope soliton more effectively than the cubic or Kerr nonlinearity (local) by altering the self-focusing or self-defocusing effects and boosting the nonlinear interactions over a wide parameter regime. In contrast, the cubic nonlinearity provides a direct nonlinear response that is crucial for stabilizing solitons. Thus, a weak nonlocal nonlinear effect (with a higher value of $\sigma$) tends to shrink the chaotic domains in the wave number space.    
	\par
\begin{figure*}[!ht]
		\centering
		\includegraphics[width=6.5in,height=2.3in]{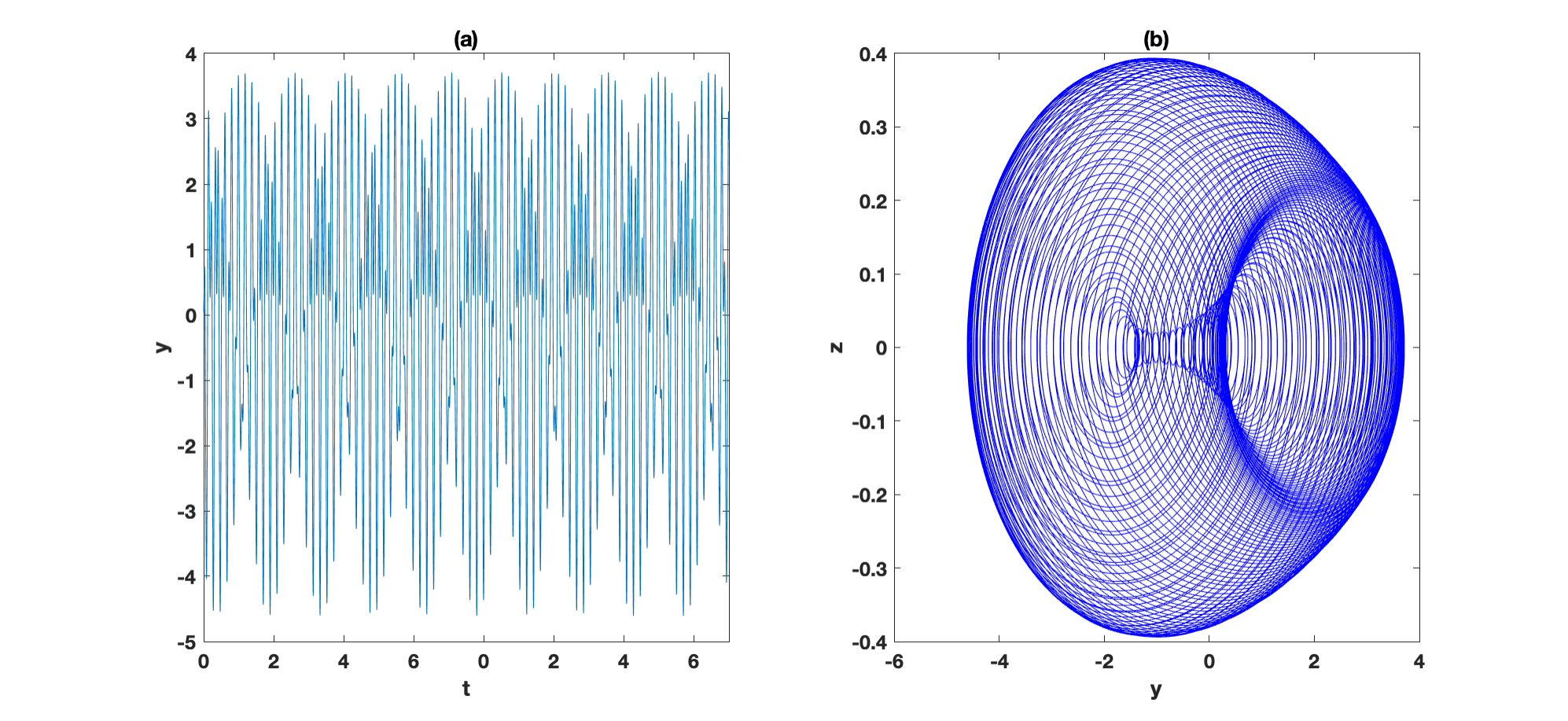}
		\caption{The time series and a quasiperiodic (torus) phase portrait are shown corresponding to the parameter values $R_0=0.8,~ k=0.35,~\sigma=0.06,~\beta_0=0.05$, and $N=0.2$.}
		\label{fig-quasi1}
	\end{figure*}

\begin{figure*}[!ht]
		\centering
		\includegraphics[width=6.5in,height=2.3in]{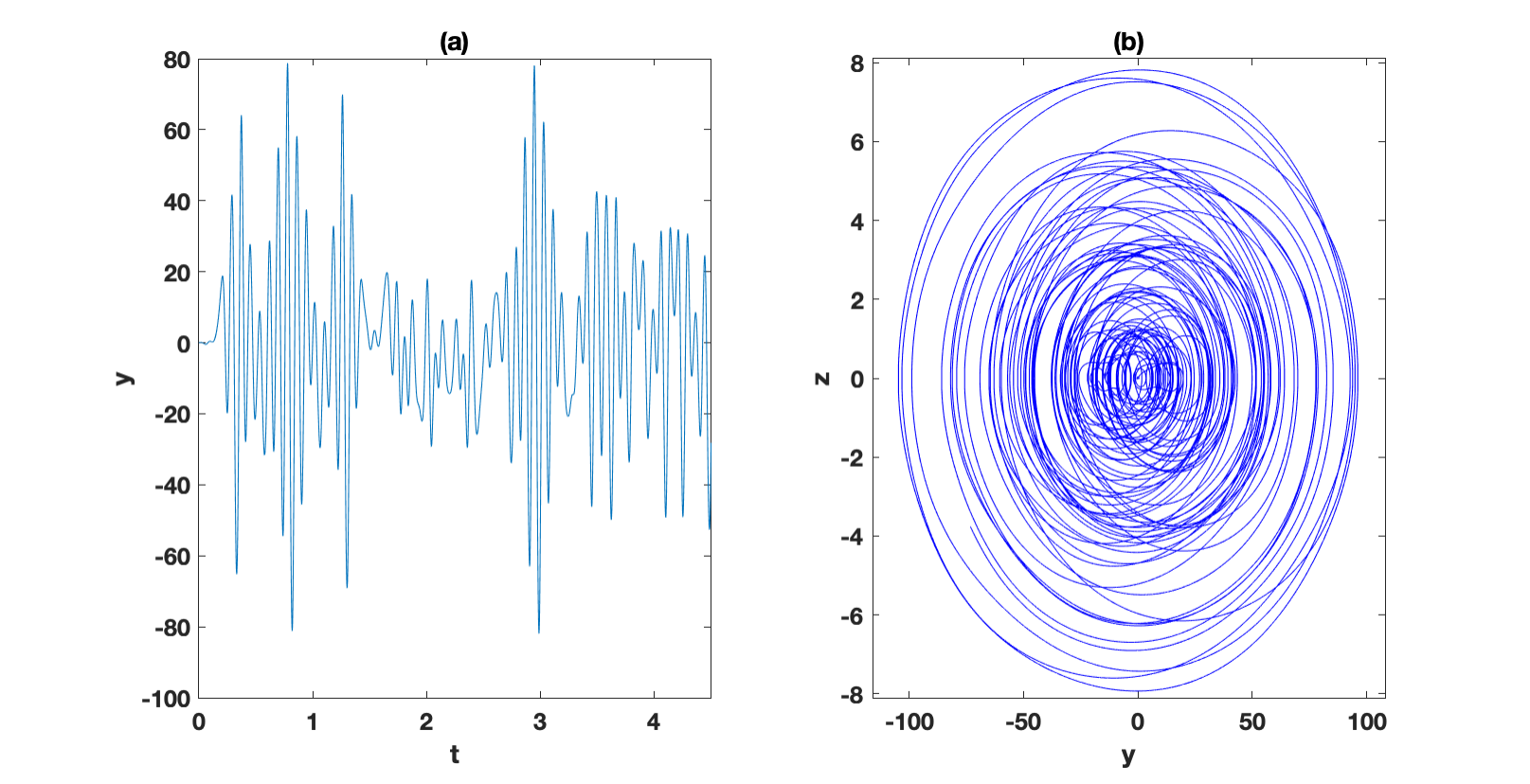}
		\caption{The time series and a chaotic attractor are shown with the same set of parameters as Fig. \ref{fig-quasi1} but for a lower value of $k$, $k=0.18$.}
		\label{fig-chaos1}
	\end{figure*}

\begin{figure*}[!ht]
		\centering
		\includegraphics[width=6.5in,height=2.3in]{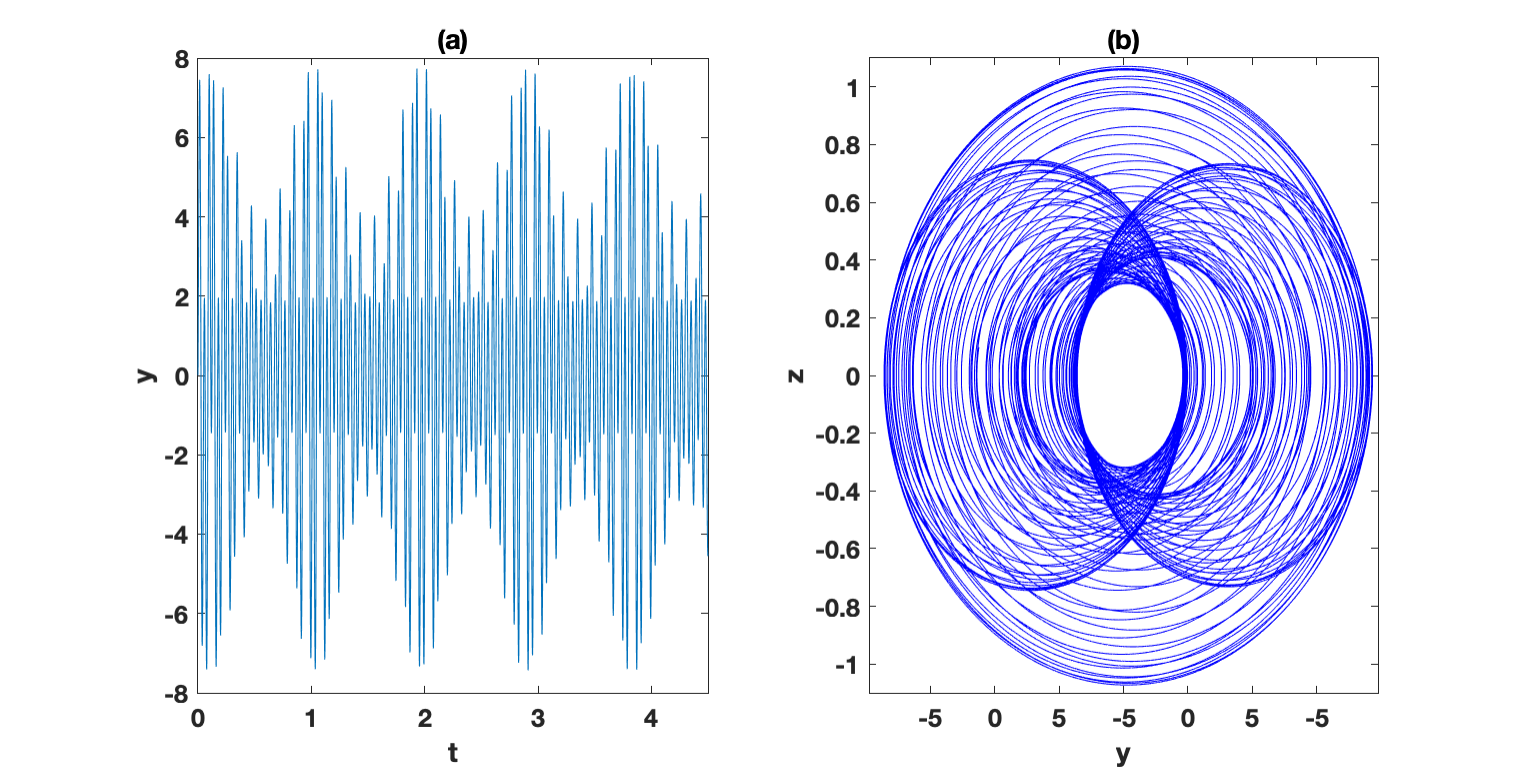}
		\caption{The same as in Fig. \ref{fig-quasi1} but for different values  of $R_0$, $k$, and $N$: $R_0=1.2$, $k=0.32$, and $N=1$.   }
		\label{fig-quasi2}    
	\end{figure*}

\begin{figure*}[!ht]
		\centering
		\includegraphics[width=6.5in,height=2.3in]{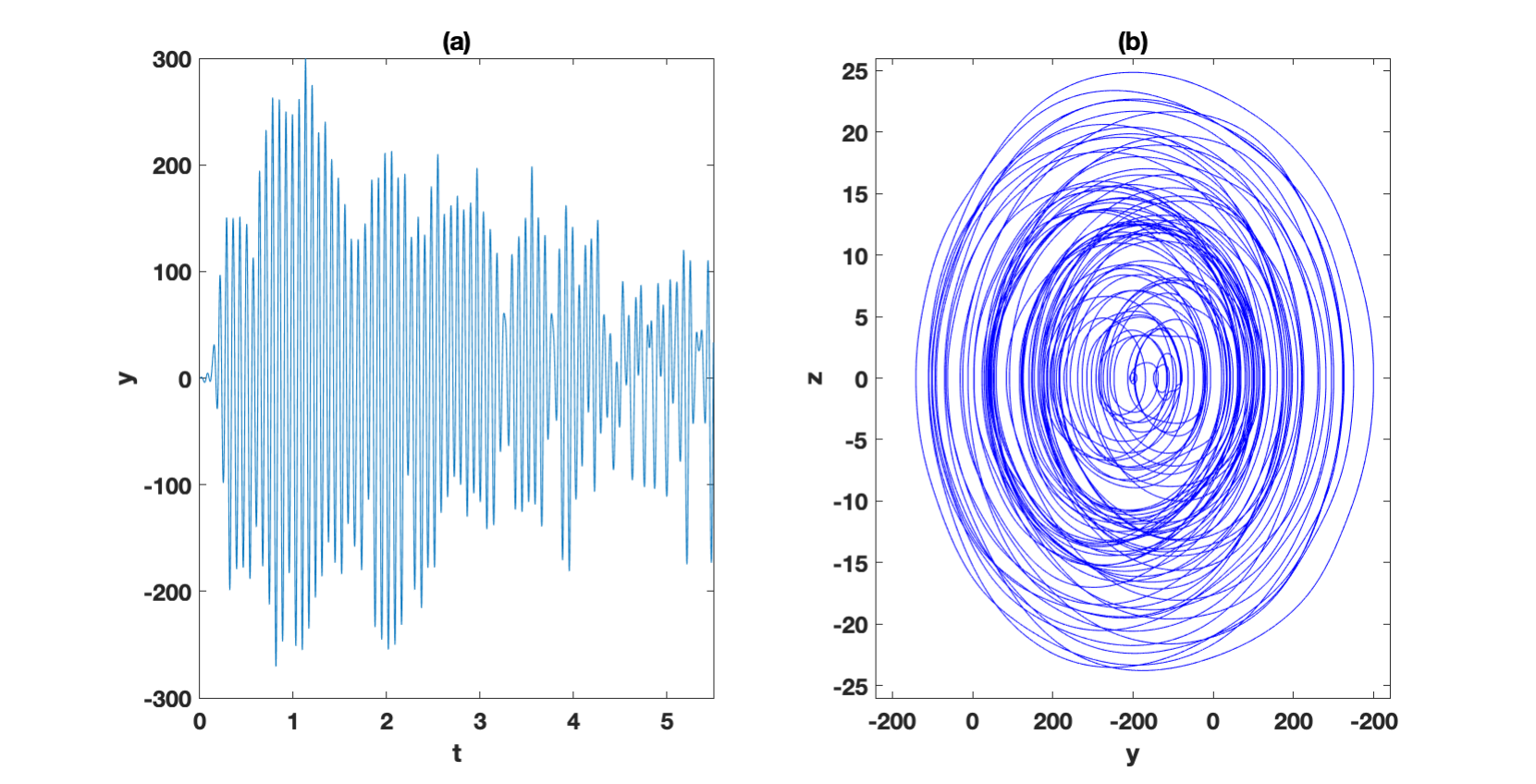}
		\caption{The same as in Fig. \ref{fig-chaos1} but for different values  of $R_0$, $k$, and $N$: $R_0=1.2$, $k=0.18$, and $N=3$.  }
		\label{fig-chaos2}
	\end{figure*}

 
	To illustrate the time series of a state variable and phase-space portraits, we numerically solve Eq. \eqref{eq-reduced} using the fourth-order Runge-Kutta scheme with a time step $dt=10^{-3}$ and consider different sets of parameter values of $R_0$ and $k$ that correspond to quasiperiodic and chaotic states (See Figs. \ref{fig-bifur-lyap1}--\ref{fig-bifur-lyap3}). We exhibit the results in Figs. \ref{fig-quasi1}--\ref{fig-chaos2}. Figures \ref{fig-quasi1} and \ref{fig-chaos1} show that keeping $R_0$ fixed in the weakly relativistic regime, as we reduce the value of $k$ from $k=0.35$ to $k=0.18$, the quasiperiodic state transits into a chaotic state. A similar phenomenon also occurs in the strong-degenerate regime [with $R_0>1$, see Figs. \ref{fig-quasi2} and \ref{fig-chaos2}]. In this context, we must note that smaller values of $k$ than $k\sim0.35$ may not be relevant to the three-wave interaction model, as they correspond to a larger number of wave modes than three. In this situation, the full system of equations \eqref{eq4} and \eqref{eq5} may be appropriate to study. However, such an investigation is beyond the scope of the present study. 
	\par
	We mention that the transitions observed in our bifurcation diagrams (Figs. \ref{fig-bifur-lyap1}-\ref{fig-bifur-lyap3}) are not numerical artifacts but also represent distinct physical regimes of wave plasma interaction. The onset of chaos occurs when the energy transfer from the high-frequency EM soliton to the low-frequency electron plasma waves becomes nonresonant. In weakly relativistic regimes ($R_0 < 1$), the plasma electrons are readily displaced by the ponderomotive force, leading to strong nonlinear feedback that destabilizes the soliton envelope. However, as the degeneracy parameter $R_0$ increases toward the strong degenerate regime ($R_0 > 1$), the relativistic degeneracy pressure provides an additional restoring force. Such an effect stiffens the plasma medium, physically limiting the amplitude of density perturbations ($n$) and thereby narrowing the subintervals where chaotic energy redistribution can occur. This mechanistic stabilizing effect of degeneracy is a key finding of our model, suggesting that superdense astrophysical plasmas are less prone to wave turbulence than their classical counterparts. 
	
	\section{Power spectrum and Poincar\'e section} 
	\par
\begin{figure*}[!ht]
		\centering
		\includegraphics[width=7in,height=4in]{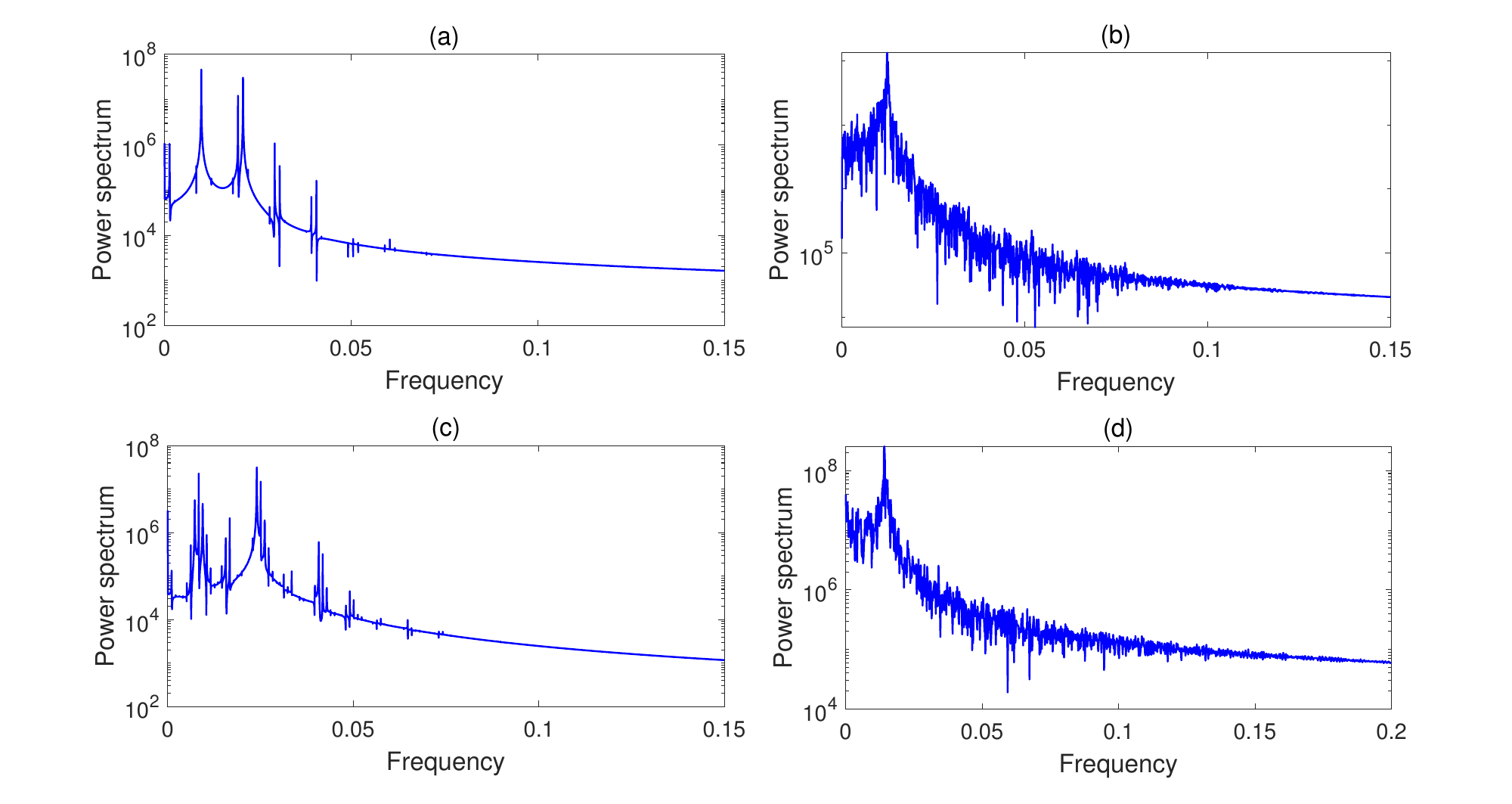}
		\caption{Power spectra for the parameter values considered in Figs. \ref{fig-quasi1}–\ref{fig-chaos2} are shown in subplots (a)–(d), respectively.}
		\label{fig-power}
	\end{figure*}

  To understand the route to chaos in the system, we analyze the evolution of the power spectrum corresponding to different dynamical regimes as the control parameters $R_0$ and $k$ are varied. The results are shown in Fig. \ref{fig-power}. For the case shown in Fig. \ref{fig-quasi1}, the power spectrum [subplot (a)] consists of several discrete sharp peaks, which is a clear signature of a quasiperiodic state. A reduction of the parameter to $k=0.18$ (Fig. \ref{fig-chaos1}) results in a broadband power spectrum [subplot (b)], which corresponds to a chaotic state. Similarly, for the case shown in Fig. \ref{fig-quasi2}, when the control parameter $R_0$ increases to $R_0=1.2$ with $k=0.32$, the power spectrum [subplot (c)] again exhibits discrete spectral peaks, indicating the appearance of quasiperiodic behavior. The power spectrum corresponding to Fig. \ref{fig-chaos2} displays a broadband continuous structure [subplot (d)], which is consistent with chaotic dynamics. The transition from quasiperiodic dynamics to chaotic behavior occurs as the parameters $R_0$ and $k$ are varied.
	\par
\begin{figure*}[!ht]
		\centering
		\includegraphics[width=5in,height=2.5in]{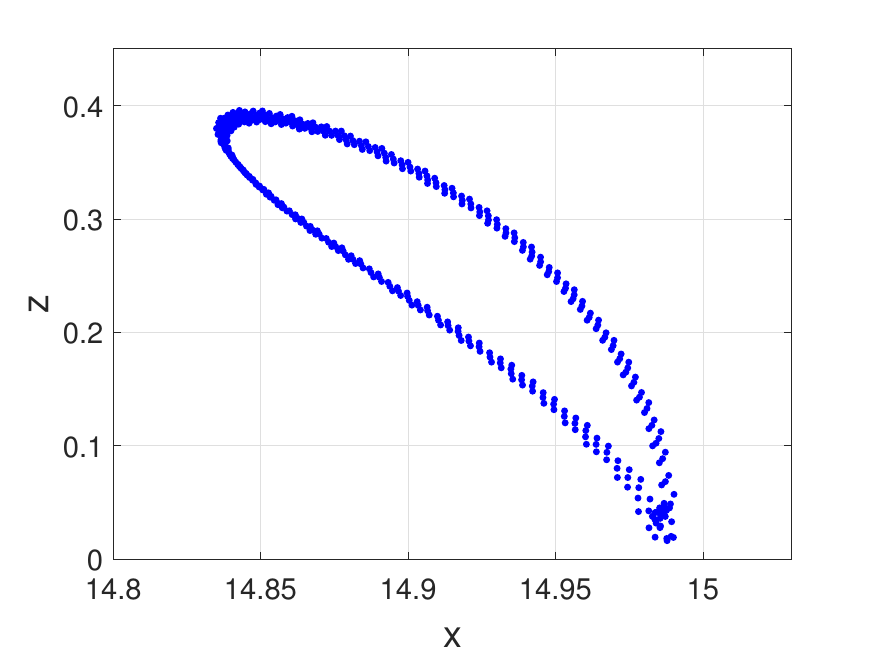}
		\caption{Poincar\'e section corresponding to the parameter set of Fig. \ref{fig-quasi1} is shown. The section is constructed by fixing $y=-0.25$ and is projected onto the $x-z$ plane.}
		\label{fig-poincare1}
	\end{figure*}

\begin{figure*}[!ht]
		\centering
		\includegraphics[width=5in,height=2.5in]{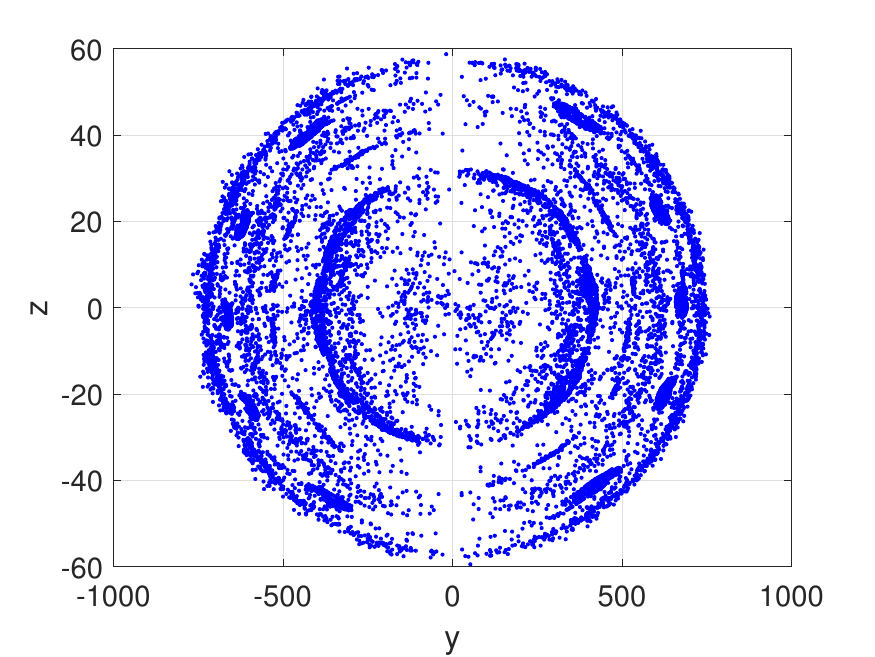}
		\caption{Poincar\'e section corresponding to the set of parameters of Fig. \ref{fig-chaos1} is shown. For this case, we fix $x=-1$ to construct $S$, and the Poincar\'e section is projected onto $y-z$ plane.}
		\label{fig-poincare2}
	\end{figure*}

\begin{figure*}[!ht]
		\centering
		\includegraphics[width=5in,height=2.5in]{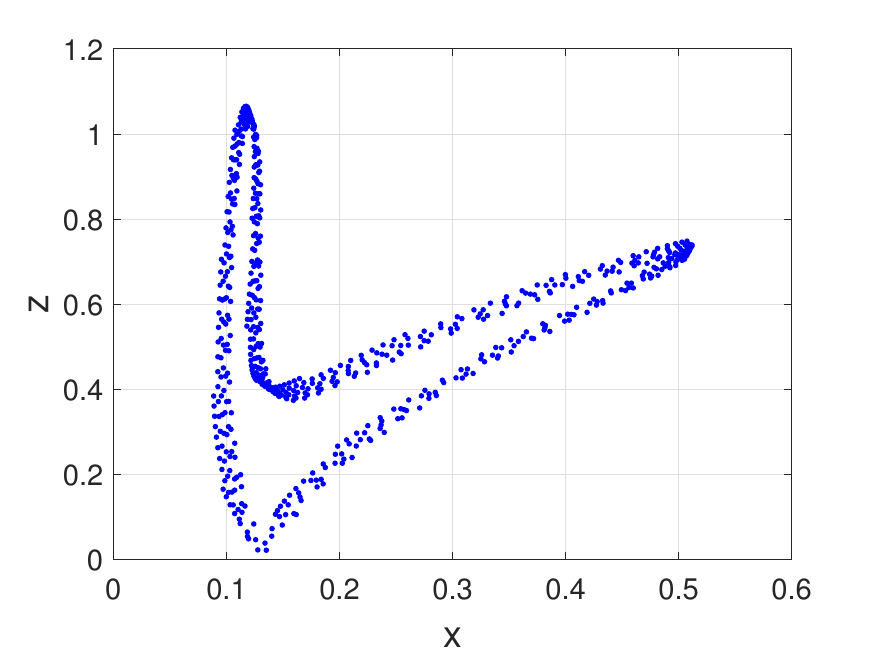}
		\caption{Poincar\'e section corresponding to the set of parameters of Fig. \ref{fig-quasi2} is shown. For this case, we fix $y=2.2$ to construct $S$, and the Poincar\'e section is projected onto $x-z$ plane.}
		\label{fig-poincare3}    
	\end{figure*}

\begin{figure*}[!ht]
		\centering
		\includegraphics[width=5in,height=2.5in]{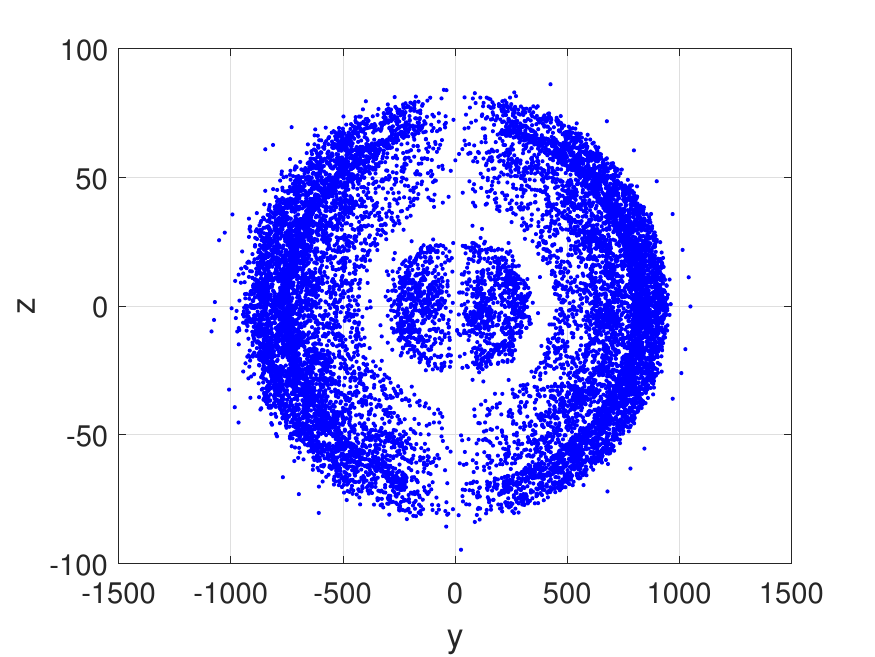}
		\caption{Poincar\'e section corresponding to the set of parameters of Fig. \ref{fig-chaos2} is shown. For this case, we fix $x=-3$ to construct $S$, and the Poincar\'e section is projected onto $y-z$ plane.}
		\label{fig-poincare4}
	\end{figure*}

 
	To analyze the system's dynamics, we employ the concept of a Poincar\'e map. For a four-dimensional dynamical system, the surface of section $S$ is three-dimensional, and we choose $S$ to be transverse to the flow. The Poincar\'e map $P$ is defined by following a trajectory starting from a point $\mathbf{x}_k \in S$ until it returns to the surface $S$ for the next time. Denoting successive intersections of the trajectory with $S$ by $\mathbf{x}_k$ and $\mathbf{x}_{k+1}$, the Poincar\'e map is given by $\mathbf{x}_{k+1} = P(\mathbf{x}_k)$. In the present study, we construct the surface of section $S$ by fixing one state variable (e.g., $x = x_0$), which defines a hyperplane in the four-dimensional phase space. The Poincar\'e section then consists of the successive intersections of the system trajectory with this surface ($S$). Since the surface $S$ is three-dimensional, one can visualize it by plotting a two-dimensional projection of the trajectory's intersections with $S$. The Poincar\'e sections corresponding to the parameter values as for Figs. \ref{fig-quasi1}–\ref{fig-chaos2} are shown in Figs. \ref{fig-poincare1}-\ref{fig-poincare4}. For the parameter set corresponding to Fig. \ref{fig-quasi1}, we fix $y = -0.25$ to construct $S$, and the Poincar\'e section projected onto the $x$–$z$ plane. The resulting Poincar\'e section forms a closed curve (Fig. \ref{fig-poincare1}), indicating a quasiperiodic behavior of the system. Corresponding to Fig. \ref{fig-chaos1}, the Poincar\'e section appears as a cloud of points, suggesting chaotic motion (Fig. \ref{fig-poincare2}). For the cases shown in Figs. \ref{fig-quasi2} and \ref{fig-chaos2}, we observe a closed curve and a scattered cloud, corresponding to quasiperiodic (Fig. \ref{fig-poincare3}) and chaotic states (Fig. \ref{fig-poincare4}), respectively.
	
		\section{Discussion and conclusion} \label{sec-discuss}
	We have studied the nonlinear interaction between high-frequency intense electromagnetic wave envelopes and low-frequency (slow plasma response) electron plasma perturbations in a relativistic degenerate plasma composed of relativistic degenerate electrons and stationary ions, and thereby proposed a spatiotemporal model of coupled nonlinear partial differential equations. We have performed a modulational instability analysis and shown that the EM wave envelope can be unstable over a finite range of modulation wave numbers. The instability growth rate gets significantly reduced by the influence of strong degeneracy effects (characterized by $R_0>1$) that correspond to high-density plasmas (such as those in the environments of white dwarfs or high-density laser-plasma interaction experiments), as well as a slight enhancement of the nonlocal parameter $\sigma$ that corresponds to a weak nonlocal nonlinear effect. To study the temporal dynamics of the model, we have considered a reduced system that governs the interaction of three waves: one unstable EM wave and two associated with low-frequency electrostatic electron density perturbations. 
	\par 
	We have studied the dynamical features of the low-dimensional reduced system. As a starting point, we performed the linear stability analysis of the system about an equilibrium point. We observed that while at least one eigenvalue of the linearized perturbed system can assume positive values, the others can be zero or negative, implying that the system can undergo quasiperiodic or chaotic states depending on the wave number of modulation $k$ and the plasma parameters $R_0$ and $\sigma$. To confirm the emergence of these states, we have performed analyses of Lyapunov exponent spectra and bifurcation diagrams, and numerically solved the temporal model to visualize the time series and phase-space portraits characterized by these analyses. In addition, to characterize the dynamical states and to visualize the phase-space topology, we have presented the power spectral density and Poincar{\'e} sections for different parameter regimes. The transition from discrete, sharp peaks (quasiperiodicity) to a broadband, continuous spectrum provides further evidence for the onset of chaos in EM solitons. Also, the Poincar{\'e} sections clearly differentiate between the closed loops of quasiperiodic tori and the scattered clouds indicative of chaotic strange attractors. 	
	 We have observed that there typically exist two domains of the modulation wave number $k$, in one of which the system exhibits chaos, and in the other, quasiperiodic states may emerge. Also, the higher the values of $R_0$ and $\sigma$, the lower the domains of the wave number for the existence of chaos. In other words, high-density degenerate plasmas or reduced nonlocal nonlinear effects favor the stability of the evolution of electromagnetic wave envelopes under the influence of electron density perturbations.
	\par
	We mention that the chaotic dynamics observed in this system arise from the fundamental nonlinearities inherent in intense wave-wave interactions. The nonlinear terms in our model (Eqs. 6 and 7) arise from the coupling between the intense circularly polarized EM wave and the slow response of electron plasma density perturbations. The key factors in the origin of the nonlinearities are: (i) the relativistic nonlinearity arises because the Lorentz factor gets modified by the intense EM wave amplitude and the electron density, and it appears as the $1/\gamma$ factor in the EM wave equation, (ii) the nonlinearity due to ponderomotive arises from the radiation pressure of the EM wave, which drives low-frequency electron density perturbations. While the parameter $\eta$ is responsible for the pondermotive nonlinearity, the parameter $\beta_0$ is for the nonlocal nonlinearity in which $\sigma$ plays the key role for the higher-order correction to the nonlocal nonlinear response in a relativistic degenerate environment. The interplay between these relativistic and ponderomotive effects, coupled with the group velocity and plasma dispersions, results in complex dynamics that lead to the emergence of quasiperiodic and chaotic states. 
	\par 
	The proposed model has several unique results that provide new insights into the dynamics of superdense plasmas. We have found that relativistic degeneracy (measured by $R_0$) serves as a stabilizing factor; as $R_0$ increases, the system's susceptibility to chaos diminishes, and the domain for stable, quasiperiodic solitons expands. This finding is significant for understanding the long duration of localized radiation bursts in white dwarf environments. Furthermore, our results show that the higher-order nonlocal correction ($\sigma$), a parameter that is intrinsic to this specific coupled framework, critically regulates the instability threshold. Specifically, we find that a weak nonlocal effect (represented by a higher value $\sigma$) effectively terminates the sensitive dependence on initial conditions, thereby preventing the onset of chaos. Traditional cubic NLS models do not capture these features, underscoring the importance of including relativistic degeneracy and nonlocal corrections in the study of high-intensity laser-plasma-like interactions. 
	\par 
	To conclude, the appearance of chaotic states in a low-dimensional model could be a signature for the emergence of spatiotemporal chaos \cite{Chow1992} and EM wave turbulence \cite{Chian1998} in the full model equations \eqref{eq4} and \eqref{eq5} that can be relevant to the environments of superdense astrophysical bodies like white dwarfs or next-generation high-density highly intense laser-plasma interaction experiments.   
	\section*{Acknowledgements} S.D.A. thanks the University Grants Commission (UGC), Government of India, for support through a junior research fellowship (JRF) with NTA reference no. 211610078837.
	\section*{Author declarations}
	\subsection*{Conflict of interest} The authors have no conflicts to disclose.
	\subsection*{Author contributions}
	Subhrajit Roy: Formal analysis (equal); Investigation (equal); Methodology (equal); Software (equal); Visualization (equal); Writing – original draft (equal). Sukhendu Das Adhikary: Formal analysis (equal); Investigation (equal); Methodology (equal); Software (equal); Visualization (equal). Amar Prasad Misra: Conceptualization (lead); Formal analysis (equal); Investigation (lead); Methodology (lead); Supervision (lead); Validation (equal); Writing – review \& editing (lead).
	\section*{Data Availability} The data that support the findings of this study are available from the corresponding author on request.
		\bibliographystyle{apsrev4-2} 
	\bibliography{reference}

@article{Zakharov1972Exact,
author = {V. E. Zakharov and A. B. Shabat},
title = {Exact Theory of Two‐Dimensional Self‐Focusing and One‐Dimensional Self‐Modulation of Waves in Nonlinear Media},
journal = {Sov. Phys. JETP},
volume = {34},
number = {1},
pages = {62–69},
year = {1972},
note = {Originally published as Zh. Eksp. Teor. Fiz. 61, 118–134 (1971)}
}

@article{Taniuti1969Modulational,
author = {T. Taniuti and N. Yajima},
title = {Modulational Instability and Envelope Solitons for Nonlinear Alfvén Waves},
journal = {J. Math. Phys.},
volume = {10},
pages = {1369},
year = {1969}
}

@article{Tidman1996,
  author={D. A. Tidman and R. A. Shanny and R. N. Sudan},
  title={Ion acoustic solitons in a weakly relativistic magnetized warm plasma},
  journal={Phys. Rev. E},
  volume={54},
  number={5},
  pages={5844--5850},
  year={1996},
  doi={10.1103/PhysRevE.54.5844}
}

@article{Markovich1998,
  author={Y. A. Markovich and N. E. Andreev and A. V. Arzhannikov},
  title={Ion motion and finite temperature effect on relativistic strong plasma waves},
  journal={Phys. Rev. E},
  volume={58},
  number={6},
  pages={7799--7807},
  year={1998},
  doi={10.1103/PhysRevE.58.7799}
}

@article{ElLabany2020,
  author={S. K. El-Labany and W. F. El-Taibany and E. E. Behery and R. Abd-Elbaki},
  title={Oblique collision of ion acoustic solitons in a relativistic degenerate plasma},
  journal={Scientific Reports},
  volume={10},
  pages={16152},
  year={2020},
  doi={10.1038/s41598-020-72449-x}
}

@article{RoyMisra2022,
  author={Sima Roy and Amar P. Misra},
  title={Electromagnetic solitons and their stability in relativistic degenerate dense plasmas with two electron species},
  journal={Frontiers in Astronomy and Space Sciences},
  volume={9},
  pages={1007584},
  year={2022},
  doi={10.3389/fspas.2022.1007584}
}

@article{Dey2024,
  author = {R. Dey and G. Banerjee and A. P. Misra and C. Bhowmik},
  title = {Ion-acoustic solitons in a relativistic Fermi plasma at finite temperature},
  journal = {Scientific Reports},
  volume = {14},
  pages = {15328},
  year = {2024},
  doi = {10.1038/s41598-024-75051-7}
}

@article{MisraBhowmik2009,
  author    = {A. P. Misra and C. Bhowmik},
  title     = {Modulational instability and envelope excitation of ion-acoustic waves in quantum electron-positron-ion plasmas},
  journal   = {Physics of Plasmas},
  volume    = {16},
  pages     = {072116},
  year      = {2009},
  doi       = {10.1063/1.3187907}
}

@article{Banerjee2010,
  title = {Spatiotemporal chaos and the dynamics of coupled Langmuir and ion-acoustic waves in plasmas},
  author = {Banerjee, S. and Misra, A. P. and Shukla, P. K. and Rondoni, L.},
  journal = {Phys. Rev. E},
  volume = {81},
  issue = {4},
  pages = {046405},
  numpages = {9},
  year = {2010},
  month = {Apr},
  publisher = {American Physical Society},
  doi = {10.1103/PhysRevE.81.046405}
 }

@article{Misra2024,
title = {Pattern formation and spatiotemporal chaos in relativistic degenerate plasmas},
journal = {Communications in Nonlinear Science and Numerical Simulation},
volume = {157},
pages = {109768},
year = {2026},
issn = {1007-5704},
doi = {https://doi.org/10.1016/j.cnsns.2026.109768},
author = {S. {Das Adhikary} and A.P. Misra}
}

@article{Kaup1978,
  author = {D. J. Kaup},
  title = {The three-wave resonant interaction},
  journal = {Rocky Mountain Journal of Mathematics},
  volume = {8},
  pages = {283--324},
  year = {1978},
  doi = {10.1216/RMJ-1978-8-1-283}
}

@article{Chian1996,
  author = {A. C.-L. Chian},
  title = {Nonlinear Wave-Wave Interactions in Astrophysical and Laboratory Plasmas},
  journal = {Astrophysics and Space Science},
  volume = {242},
  pages = {249},
  year = {1996}
}

@article{Rososhek2025,
  author = {A. Rososhek and E. S. Lavine and B. R. Kusse and W. M. Potter and D. A. Hammer},
  title = {Randomization of a laser wave front by the turbulent gas-puff Z-pinch plasma column},
  journal = {Phys. Rev. E},
  volume = {111},
  pages = {045202},
  year = {2025},
  doi = {10.1103/PhysRevE.111.045202}
}

@article{Zulick2025,
  author = {C. Zulick and L. Willingale and A. G. R. Thomas and P. M. Nilson and A. Maksimchuk and W. Nazarov and T. C. Sangster and C. Stoeckl and K. Krushelnick},
  title = {Measurement of field generation at plasma interfaces due to fast electron beam currents from high-intensity laser–plasma interactions},
  journal = {Physics of Plasmas},
  volume = {32},
  number = {7},
  pages = {072113},
  year = {2025}
}

@article{Yuan2025,
  author = {C. Yuan and Vitor Cardoso and Francisco Duque and and Ziri Younsi},
  title = {Gravitational waves from accretion disks: Turbulence and emission in compact objects},
  journal = {arXiv preprint},
  year = {2025},
  eprint = {2502.07871},
  archivePrefix = {arXiv},
  primaryClass = {gr-qc}
}

@article{Beattie2025,
  author = {James Beattie and  Christoph Federrath and S. Klessen and Salvatore Cielo and Amitava Bhattacharjee},
  title = {The spectrum of magnetized turbulence in the interstellar medium},
  journal = {Nature Astronomy},
  volume={9},
  year = {2025},
  pages = {1195--1205}
}

@article{Holkundkar2018,
  title = {Transition from wakefield generation to soliton formation},
  author = {Holkundkar, Amol R. and Brodin, Gert},
  journal = {Phys. Rev. E},
  volume = {97},
  issue = {4},
  pages = {043204},
  numpages = {6},
  year = {2018},
  month = {Apr},
  publisher = {American Physical Society},
  doi = {10.1103/PhysRevE.97.043204}
  }

@article{misra2018,
    author = {Misra, A. P. and Chatterjee, Debjani},
    title = {Stimulated scattering instability in a relativistic plasma},
    journal = {Physics of Plasmas},
    volume = {25},
    number = {6},
    pages = {062116},
    year = {2018},
    month = {06},
    issn = {1070-664X},
    doi = {10.1063/1.5037955},
    url = {https://doi.org/10.1063/1.5037955}
   }

@article{misrapla2008,
title = {A novel hyperchaos in the quantum Zakharov system for plasmas},
journal = {Physics Letters A},
volume = {372},
number = {9},
pages = {1469},
year = {2008},
issn = {0375-9601},
doi = {https://doi.org/10.1016/j.physleta.2007.09.054},
author = {A.P. Misra and D. Ghosh and A.R. Chowdhury}
}

@article{Sharma2005,
    author = {Sharma, R. P. and Batra, K. and Verga, A. D.},
    title = {Nonlinear evolution of the modulational instability and chaos using one-dimensional Zakharov equations and a simplified model},
    journal = {Physics of Plasmas},
    volume = {12},
    number = {2},
    pages = {022311},
    year = {2005},
    month = {01},
    issn = {1070-664X},
    doi = {10.1063/1.1850477},
    url = {https://doi.org/10.1063/1.1850477}
    }

@article{sroy2023chaos,
    author = {Roy, Subhrajit and Roy, Animesh and Misra, Amar P.},
    title = {Chaos and complexity in the dynamics of nonlinear Alfvén waves in a magnetoplasma},
    journal = {Chaos: An Interdisciplinary Journal of Nonlinear Science},
    volume = {33},
    number = {2},
    pages = {023130},
    year = {2023},
    month = {02},
    issn = {1054-1500},
    doi = {10.1063/5.0138866},
    url = {https://doi.org/10.1063/5.0138866}
   }

@article{roy2020,
doi = {10.1088/1402-4896/ab447d},
year = {2019},
month = {dec},
publisher = {IOP Publishing},
volume = {95},
number = {1},
pages = {015603},
author = {Roy, Sima and Chatterjee, Debjani and Misra, A P},
title = {Generation of wakefields and electromagnetic solitons in relativistic degenerate plasmas},
journal = {Physica Scripta}
}

@article{misra2010,
    author = {Misra, A. P. and Banerjee, S. and Haas, F. and Shukla, P. K. and Assis, L. P. G.},
    title = {Temporal dynamics in the one-dimensional quantum Zakharov equations for plasmas},
    journal = {Physics of Plasmas},
    volume = {17},
    number = {3},
    pages = {032307},
    year = {2010},
    month = {03},
    issn = {1070-664X},
    doi = {10.1063/1.3356059},
    url = {https://doi.org/10.1063/1.3356059}
    }

@article{Chian1998,
doi = {10.1086/306214},
year = {1998},
month = {oct},
publisher = {},
volume = {505},
number = {2},
pages = {993},
author = {Chian, Abraham C.-L. and Borotto, Félix A. and Gonzalez, Walter D.},
title = {Alfvén Intermittent Turbulence Driven by Temporal Chaos},
journal = {The Astrophysical Journal}
}

@article{Chow1992,
  title = {Spatiotemporal chaos in the nonlinear three-wave interaction},
  author = {Chow, Carson C. and Bers, A. and Ram, A. K.},
  journal = {Phys. Rev. Lett.},
  volume = {68},
  issue = {23},
  pages = {3379--3382},
  numpages = {0},
  year = {1992},
  month = {Jun},
  publisher = {American Physical Society},
  doi = {10.1103/PhysRevLett.68.3379}
  }

\end{document}